\documentclass[preprint2]{aastex}

\usepackage{graphicx,epstopdf,amsmath}
\usepackage{import}

\usepackage{color}

\newcommand{\newa}{NewA}
\newcommand{\jkas}{JKAS}

\newcommand{\phpl}{PhPl}
\newcommand{\sci}{Sci}
\newcommand{\jcoph}{JCoPh}
\newcommand{\cophc}{CoPhC}

\newcommand{\soph}{SoPh}
\newcommand{\phfl}{PhFl}
\newcommand{\jgra}{JGRA}
\newcommand{\rmp}{RvMP}
\newcommand{\jetp}{JETP}
\newcommand{\nature}{Natur}
\newcommand{\rspsa}{RSPSA}

\begin{document}

\title{Three-Dimensional Simulations of Tearing and Intermittency in Coronal Jets}

\author{P.~F.~Wyper} 
\affil{Universities Space Research Association, NASA Goddard Space Flight Center, 8800 Greenbelt Rd, Greenbelt, MD 20771}
\email{peter.f.wyper@nasa.gov}

\author{C.~R.~DeVore} 
\affil{Heliophysics Science Division, NASA Goddard Space Flight Center, 8800 Greenbelt Rd, Greenbelt, MD 20771}
\email{c.richard.devore@nasa.gov}

\author{J.~T.~Karpen} 
\affil{Heliophysics Science Division, NASA Goddard Space Flight Center, 8800 Greenbelt Rd, Greenbelt, MD 20771}
\email{judy.karpen@nasa.gov}

\author{B.~J.~Lynch} 
\affil{Space Sciences Laboratory, University of California, Berkeley, CA 94720, USA}
\email{blynch@ssl.berkeley.edu}

\begin{abstract}
Observations of coronal jets increasingly suggest that local fragmentation and intermittency play an important role in the dynamics of these events. In this work we investigate this fragmentation in high-resolution simulations of jets in the closed-field corona. We study two realizations of the embedded-bipole model, whereby impulsive helical outflows are driven by reconnection between twisted and untwisted field across the domed fan plane of a magnetic null. We find that the reconnection region fragments following the onset of a tearing-like instability, producing multiple magnetic null points and flux-rope structures within the current layer. The flux ropes formed within the weak-field region in the center of the current layer are associated with ``blobs" of density enhancement that become filamentary threads as the flux ropes are ejected from the layer, whereupon new flux ropes form behind them. This repeated formation and ejection of flux ropes provides a natural explanation for the intermittent outflows, bright blobs of emission, and filamentary structure observed in some jets. Additional observational signatures of this process are discussed. Essentially all jet models invoke reconnection between regions of locally closed and locally open field as the jet-generation mechanism. Therefore, we suggest that this repeated tearing process should occur at the separatrix surface between the two flux systems in all jets. A schematic picture of tearing-mediated jet reconnection in three dimensions is outlined. 
\end{abstract}


\section{Introduction}
Observations show the solar atmosphere to be highly dynamic, with impulsive, energetic events occurring over a broad range of spatial and temporal scales. Magnetic reconnection, the process whereby stored magnetic energy is released via a reconfiguration of the magnetic connectivity, is generally believed to be central to the majority of such events \citep{PriestForbes2000}. In recent years, our perceived understanding of how magnetic reconnection proceeds in the corona has shifted away from the idea that reconnection occurs smoothly in a single, well defined current layer \citep[e.g.][]{Parker1957,Sweet1958,Petschek1964} towards a picture of more intermittent, fragmented dynamics involving multiple current layers and energy release sites \citep[e.g.][]{Huang2013}.

Observationally, a growing number of cases exhibit such intermittency amongst the largest and best-resolved events. Bright blobs are observed in the ray-like features that form beneath erupting coronal mass ejections (CMEs) when viewed along their axes \citep[e.g.][]{Lin2005,Guo2013}. Dark, void-like supra-arcade downflows \citep{McKenzie1999,McKenzie2009} are observed in post-CME {rays} when viewed from the {side}. Additionally, radio pulsations \citep{Kliem2000}, plasma blobs \citep[e.g.][]{Ohyama1998}, and wave-like motions of the flare ribbons \citep{Brannon2015} suggest that bursty reconnection occurs in solar flares. Intermittent plasma outflows and blobs have also been observed in filament eruptions \citep{Reeves2015}. {All of these features suggest an intermittent, bursty reconnection process.}

The onset and nonlinear evolution of the tearing instability \citep{Furth1963} provides a natural explanation for much of this fragmentation and intermittent reconnection. Indeed, tearing and the associated formation of magnetic islands/flux ropes have been observed in numerical simulations of CMEs and flares \citep{Barta2011,Karpen2012,Guo2013b,Lynch2013} and surges \citep{Karpen1996}, as well as during more gentle quasi-steady interchange reconnection \citep{Edmondson2010}. In a self-consistently evolving system, where current layers form dynamically over time, tearing is initiated when a stable {current} layer becomes sufficiently long and thin. In two dimensions, numerical studies show that this typically occurs when $S = Lv_{a}/\eta > 10^{4}$, where $S$ is the Lundquist number based on the length $L$ of the current layer, $v_{a}$ is the inflow Alfv\'{e}n speed, and $\eta$ is the plasma resistivity \citep[e.g.][]{Biskamp1986}. In the context of such lengthening and thinning current layers, the tearing instability is typically referred to as the ``plasmoid instability.'' \citet{Loureiro2007} were the first to develop a two-dimensional linear theory describing how this instability grows in a pre-existing Sweet-Parker sheet. Subsequently, \citet{Pucci2014} argued that such a sheet is unattainable in nature, and that any developing current layer will disrupt before it reaches the aspect ratio consistent with the Sweet-Parker scaling. Regardless of the exact nature of the linear phase, if the global evolution is sufficiently slow, the subsequent nonlinear dynamics will be dominated by the formation, coalescence and ejection of magnetic islands. In the corona, $S$ is orders of magnitude higher than $10^{4}$, and long, thin current layers are expected to form in non-potential magnetic fields on the basis of ideal modeling \citep[e.g.][]{Syrovatskii1971,Longcope1996}. Consequently, tearing-mediated reconnection appears to be inevitable.

In this work, we explore the role of tearing and the formation of fine structure in closed-field coronal jets. Coronal jets are transient, impulsive, collimated plasma outflows originating from bright regions low in the solar atmosphere.  They are smaller than typical flares or CMEs, but {can} share some characteristic features \citep[e.g.][]{Shibata1997}. The most energetic jets are observed in X-rays \citep[e.g.][]{Cirtain2007,Shimojo1996}, but jets are also observed at EUV and optical wavelengths \citep[e.g.][]{Filippov2013,Savcheva2007,Guo2013}. Typically, a brightening of the base occurs first, followed by rapid, often supersonic, plasma outflows guided by the ambient field. The morphological appearance of {the source region of} many jets is that of a sea anemone \citep{Shibata1994}, with the outflows forming a bright spire extending from a compact quasi-circular base. A large fraction of coronal jets also exhibit a helical structure to their outflows and a wandering of the jet spire when viewed against the plane of the sky \citep{Patsourakos2008}. X-ray and EUV jets are observed prolifically in coronal holes \citep[e.g.][]{Cirtain2007,Savcheva2007}, where the ambient field is quasi-unidirectional and the jets appear as extended radial spires, sometimes extending out far into the heliosphere \citep[e.g.][]{Patsourakos2008,Filippov2013}. Such jets are also observed (although less readily against the brighter background plasma) in closed-field regions, particularly near active regions \citep[e.g.][]{Torok2009,Yang2012,Guo2013,Lee2013,Schmieder2013,Zheng2013,Cheung2015}. In these cases, the jet material propagates along the ambient coronal loops and the spire often has a curved appearance. Closed-field jets also have been associated with brightening at the distant footpoint of the connecting coronal loop \citep[e.g.][]{Torok2009,Zhang2013}.

As in flares and CMEs, there is some observational evidence that intermittent, fragmented reconnection plays a role in jets. Recent {\it{Interface Region Imaging Spectrograph}} (IRIS) observations of an active-region jet revealed filamentary fine-scale structure in the emission from the reconnection region \citep{Cheung2015}. {\it{Solar Dynamics Observatory}} observations have shown blobs forming in both small \citep{Zhang2014} and large \citep{Filippov2015} open-field EUV jets. Recent {\it{Solar TErrestrial RElations Observatory}} observations have also revealed trains of plasma blobs within jets in the closed-field corona near active regions \citep{Zhang2016}. Using a new imaging technique, \citet{Chen2013} analyzed the moving sources of type III radio bursts in an active-region jet. They found multiple reconnection sites within the jet region and filamentary structures in the jet outflow. Jets originating from the cooler solar chromosphere have also been reported to contain plasma blobs and to have a multi-treaded structure \citep{Singh2011,Singh2012}. Finally, the formation of islands/flux ropes has also been reported in several jet simulations \citep{Yokoyama1994,Karpen1995,Yokoyama1996,Moreno-Insertis2013,Yang2013}.

The magnetic field associated with these events consists of a parasitic polarity patch, with a field component normal to the photosphere of one sign, embedded within a region of weaker field of the opposite sign. The field of the parasitic polarity closes down to the photosphere and is separated from the background, locally open field by a dome-shaped separatrix surface, topped with a three-dimensional (3D) magnetic null point. In open-field jets, the background field connects to the distant heliosphere, whereas in closed-field jets, the field closes back to the photosphere at a distant footpoint. This domed configuration can form as a result of flux emergence \citep[e.g.][]{Torok2009}, or be pre-existing {\citep[e.g.][]{Zhang2012,Cheung2015}}. {Such a two-flux system readily allows the displacement of the field lines near the null, forming a strong, fully 3D current {layer} that eventually will begin to reconnect} \citep{Antiochos1996,PriestTitov1996,Pontin2007}.

{If the current layers formed in jet source regions become sufficiently long and thin, it is to be expected that they will become highly fragmented following the onset of a tearing-like instability, in a manner similar to 2D Sweet-Parker-like layers.} Indeed, \citet{Wyper2014a} studied the stability of the current layers formed self-consistently at 3D null points (as occurs in solar jets) through external boundary driving. They found that rapid tearing does occur beyond a critical Lundquist number $S_{c} \approx 2\times 10^{4}$, indicating that current layers in coronal jets should be highly unstable to tearing. However, in contrast to 2D studies, the nonlinear dynamics were dominated by the complex interplay of multiple flux ropes and null points within a nearly turbulent current layer \citep{Wyper2014b}. Additionally, the finite extent of the 3D null current layer allowed twist and mass within the flux ropes to escape {in the direction perpendicular to the two outflow jets,} so that the ropes rarely grew significantly wider than the thickness of the main layer. {Using static models for the domed anemone field of coronal jets, \citet{Pontin2015} demonstrated that such flux ropes also {create structure} in the open/closed boundary.}

The aim of this work is to explore the occurrence of such fully 3D tearing in coronal jets and the role that it plays in the jetting behavior. The paper is structured as follows. In $\S\S 2$ and $3$ we introduce the model and the numerical setup. In $\S 4$ we summarize the overall evolution of two jets chosen for study, whilst in $\S 5$ we investigate the tearing-driven dynamics and fine structure during each jet. We summarize and discuss our findings in $\S 6$.

\begin{figure}[t]
\centering
\includegraphics[width=0.45\textwidth]{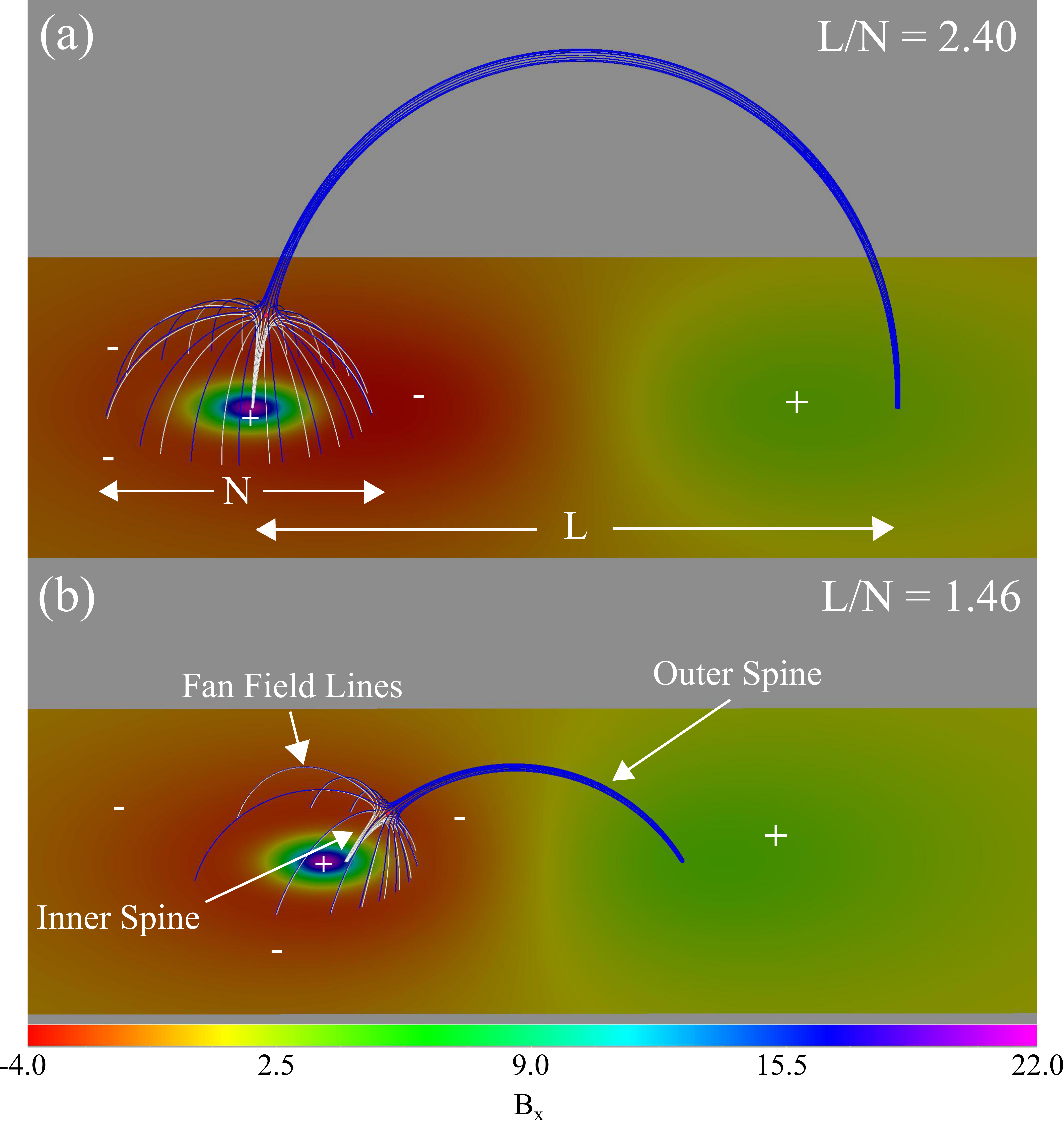}
\caption{Initial magnetic field in two configurations with aspect ratios $L/N=2.40$ (top) and $L/N=1.46$ (bottom). The bottom planes are color-shaded according to the sign ($+,-$) and strength of the field component normal to the surface ($B_x$). Selected field lines outline the fan separatrix surface and the inner and outer spine lines emanating from the magnetic null point (red sphere).}
\label{fig:fields}
\end{figure}

\section{Jet Model}
{A prototypical model for coronal jets that captures the impulsive and helical nature of many jets was presented by \citet{Shibata1986}. Subsequently, the model was refined and {explored} by \citet{Pariat2009,Pariat2010,Pariat2015}}. Twist is introduced to the magnetic field beneath a pre-existing null dome, which is embedded in an open field with a small inclination angle to the photosphere. Fast reconnection is inhibited until the onset of a kink-like instability {breaks the symmetry of the twisted field}, enabling rapid reconnection to occur. The sequential fast reconnection of magnetic flux through the null region generates a nonlinear Alfv\'{e}n wave pulse, which carries magnetic twist and plasma material out along the ambient field as an untwisting helical jet.

Recently, we applied this model to closed-field configurations and produced helical jets consistent in their energy release and morphology with jets observed in both active regions and the quiet Sun \citep[][referred to as WD16 hereafter]{Wyper2016}. We considered the simplest possible jet-generating magnetic configuration: a small-scale, strong photospheric patch of one polarity embedded in the opposite polarity of a large-scale, weaker background field that forms a closed loop. Two intrinsic length scales characterize such a system; the diameter ($N$) of the separatrix dome and the separation ($L$) of the two spine footpoints. The aspect ratio $L/N$ then quantifies the relative sizes of the anemone region and the connecting coronal loop. In terms of this ratio, open-field jets correspond to the limit $L/N \to \infty$. {It is expected that jets can occur across the full range of $L/N$. Reported values range from 2.5 to 4 in large-scale jets \citep[e.g.][]{Sun2013,Cheung2015} to $5$ and higher in small-scale jets \citep[e.g.][]{Zhang2013,Zhang2016}. Over the parameter range studied in WD16,} the threshold for triggering the kink instability that led to the jet, as well as the subsequent jet dynamics, were found to depend strongly on the ratio $L/N$. When $L$ and $N$ were comparable, the high local inclination angle led to quasi-steady slow reconnection prior to a short-lived and weak jet. For $L$ much larger than $N$, the dynamics were similar to the open-field case, with negligible reconnection prior to longer duration, more energetic jets. 

In this work, we focus on the role of tearing in two cases with $L/N = 1.46$ and $2.40$. They are near the extremes of the parameter range investigated in WD16 and illustrate these markedly different behaviors. Figure \ref{fig:fields} shows the magnetic topology of the two configurations.

\begin{figure*}[t]
\centering
\includegraphics[width=0.9\textwidth]{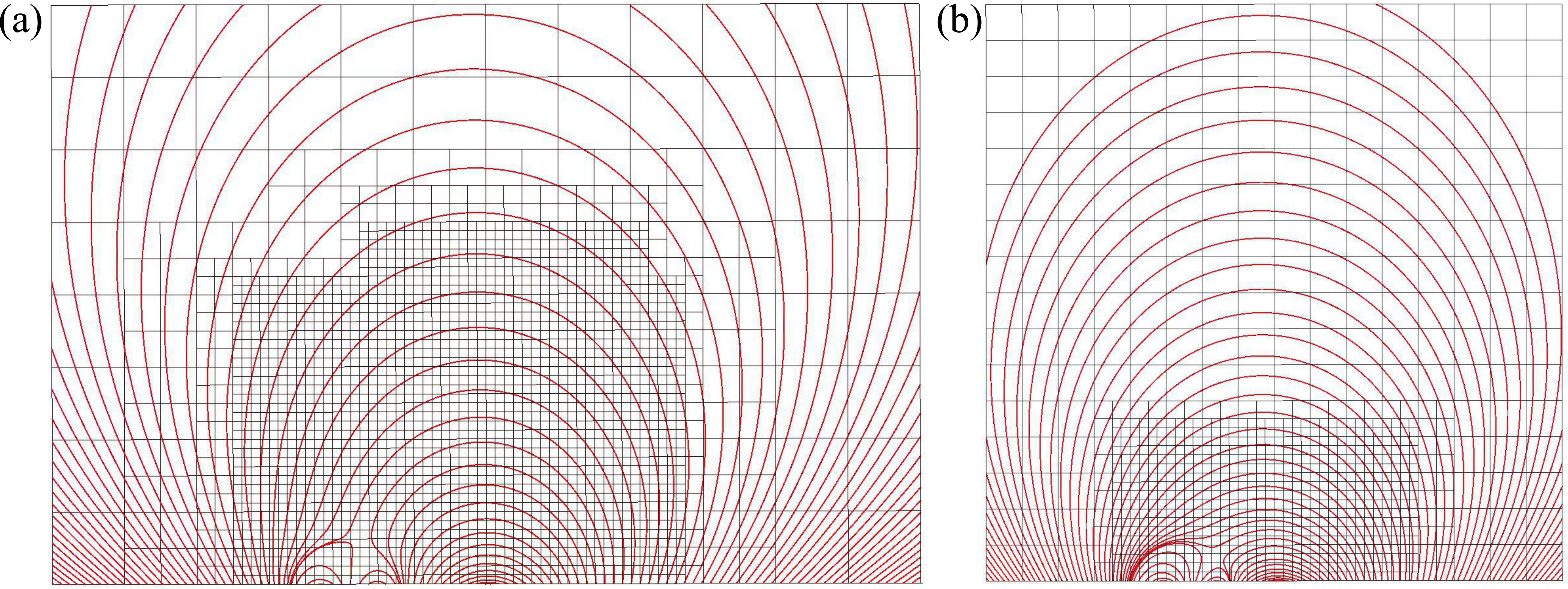}
\caption{Side view of the initial {grid lines (black) and field lines (red)} in the mid-plane of each configuration shown in Figure \ref{fig:fields}. Every block in the figure contains $8 \times 8 \times 8$ grid cells. (a) $L/N = 2.40$, (b) $L/N=1.6$.}
\label{fig:grid}
\end{figure*}

\section{Numerical Setup}
As before we use the Adaptively Refined Magnetohydrodynamics Solver \citep[ARMS;][]{DeVore2008} to solve the ideal MHD equations in the form
\begin{gather}
\frac{\partial \rho}{\partial t} + \boldsymbol{\nabla}\cdot(\rho \mathbf{v}) = 0, \\
\frac{\partial (\rho \mathbf{v})}{\partial t}+\boldsymbol{\nabla}\cdot(\rho \mathbf{v}\mathbf{v}) + \boldsymbol{\nabla} P -\frac{(\boldsymbol{\nabla}\times \mathbf{B})\times\mathbf{B}}{\mu_{0}} = 0, \\
\frac{\partial U}{\partial t}+\boldsymbol{\nabla}\cdot (U\mathbf{v})+P\boldsymbol{\nabla}\cdot\mathbf{v} = 0,\\
\frac{\partial \mathbf{B}}{\partial t}-\boldsymbol{\nabla}\times (\mathbf{v}\times\mathbf{B})=0.
\end{gather}
where $t$ is the time, $\rho$ is the mass density, $P = \rho R T$ is the thermal pressure, $U = P/(\gamma-1)$ is the internal energy density, $\mu_{0}=4\pi$ is the magnetic permeability, and $\mathbf{B}$ and $\mathbf{v}$ are the 3D magnetic and velocity fields. The equations are solved in non-dimensional form, and can be scaled to typical coronal parameters representative of either active-region or quiet-Sun jets (WD16).

The plasma pressure, temperature, and density are initially uniform in space with values of $0.1$, $1.0$ and $1.0$ respectively; the gas constant $R=0.1$. The peak vertical magnetic field at the center of the parasitic polarity in each simulation is $\vert B \vert \approx 21$, corresponding to a minimum plasma beta in the volume of $\beta \approx 6 \times 10^{-6}$, whilst away from the parasitic polarity the background weak dipole field has a minimum plasma beta of $\beta \approx 1.5\times 10^{-2}$ on the photosphere at the center of each domain ($[x,y,z]=[0,0,0]$). The background sound speed is $v_{s} \approx 0.13$, whilst the maximum Alfv\'{e}n speed $v_{a} \approx 5.9$ at the center of each parasitic polarity. The width of the separatrix ($N$) is $5.82$ and $6.34$ for $L/N = 1.46$ and $2.40$ respectively.  Thus, each time unit corresponds to roughly the Alfv\'{e}n travel time across the separatrix based on the peak Alfv\'{e}n speed at the center of the parasitic polarity, $t \approx N/v_{a} \approx 1$.

{The adaptive grid employed by ARMS is constructed from a basis set of root blocks (containing $8\times 8\times 8$ grid cells), which can be subdivided to attain higher grid refinements in a pre-defined way and/or adaptively as the solution requires \citep{MacNeice2000}. In WD16, the root blocks were $17 \times 17 \times 17$ in extent and a floor value of four levels of grid refinement was set. The grid could refine a further two times depending upon the formation of strong current layers and fine-scale structure. Additionally, a small volume covering the footprint of the separatrix surface on the photosphere was held fixed at the maximum six levels of refinement throughout each simulation, to resolve the boundary driving adequately. The grid adaptation in the main volume focused on currents forming in weak-field regions, and so resolved well the region around the null. However, the weaker fine structure generated in the coronal loop by the jet, as well as along the periphery of the current layer on the separatrix surface, were not as well resolved. In order to resolve all of this fine-scale structure uniformly for the present study, we repeated both calculations holding the volume within which each jet was confined by the connecting coronal loops at a fixed resolution using six levels of refinement. Employing a fixed, finer grid also avoided the possibility that tearing was initiated in the simulations by dynamic changes in the grid. Figure \ref{fig:grid} shows a side view of the grids that we used in the new simulations. A larger numerical box with dimensions $[0,34]\times[-25.5,25.5]\times[-17.0,17.0]$ ($2 \times 3 \times 2$ root blocks), along with a lower floor value of three refinement levels to reduce memory usage, was adopted for the $L/N=2.40$ simulation. The $L/N = 1.46$ simulation had the same dimensions as in WD16: $[0,34]\times[-17.0,17.0]\times[-8.5,8.5]$ ($2 \times 2 \times 1$ root blocks). The grid separation at the sixth level of refinement was the same in both simulations ($dl \approx 0.066$).}

The dissipation in the model is provided by a small but finite numerical diffusion that scales quadratically with the grid spacing. The new grid represents a reduction in dissipation in the affected volume by a factor of $\approx 16$, assuming similar local conditions. We found this to be sufficient resolution for rapid tearing to occur in the current layers of the jets during their evolution. This indicates that the effective Lundquist number of the current layers $S_{eff} = v_{a}L/\eta_{eff}$ (where $\eta_{eff}$ is the effective resistivity, $L$ is the sheet length in the plane of spine-fan collapse, and $v_{a}$ is the inflow Alfv\'en speed) was high enough to exceed the critical threshold for tearing in high-aspect-ratio current layers formed at 3D nulls ($S_{c} \approx 2\times 10^{4}$) identified by \citet{Wyper2014a}. Because the growth rate of the tearing instability in Sweet-Parker-like, high-aspect-ratio current layers increases with (or becomes independent of) the Lundquist number \citep{Loureiro2007,Pucci2014}, it is expected that at higher resolutions and effective Lundquist numbers the jets produced will be more unstable to tearing and the formation of fine-scale structure. As such, these simulations represent a lower bound for the complexity and dynamics expected from tearing-mitigated reconnection in coronal jets. 

The photosphere in each calculation is located at $x=0$. Free energy is injected into the system by bodily rotating each parasitic polarity via a prescribed velocity pattern on this surface, described in detail in WD16. The driving is parallel to the photosphere,  preserves the normal component of the magnetic field across this surface, and is subsonic and sub-Alfv\'{e}nic so that the field evolves quasi-statically prior to each jet. The driving is ramped up from and back down to zero using a cosine profile over a period of $1000$ Alfv\'{e}n times, after which no further driving occurs. All boundaries are closed (zero fluxes of mass, momentum, energy and magnetic flux pass through). Free slip conditions are imposed on the top and side boundaries, whereas the bottom boundary is line-tied with zero tangential velocity except within the driving region where the flow is prescribed. Each calculation was run out to $t=1200$, long enough for the jets to occur and the system to begin to relax.

\begin{figure*}[t]
\centering
\includegraphics[width=1.0\textwidth]{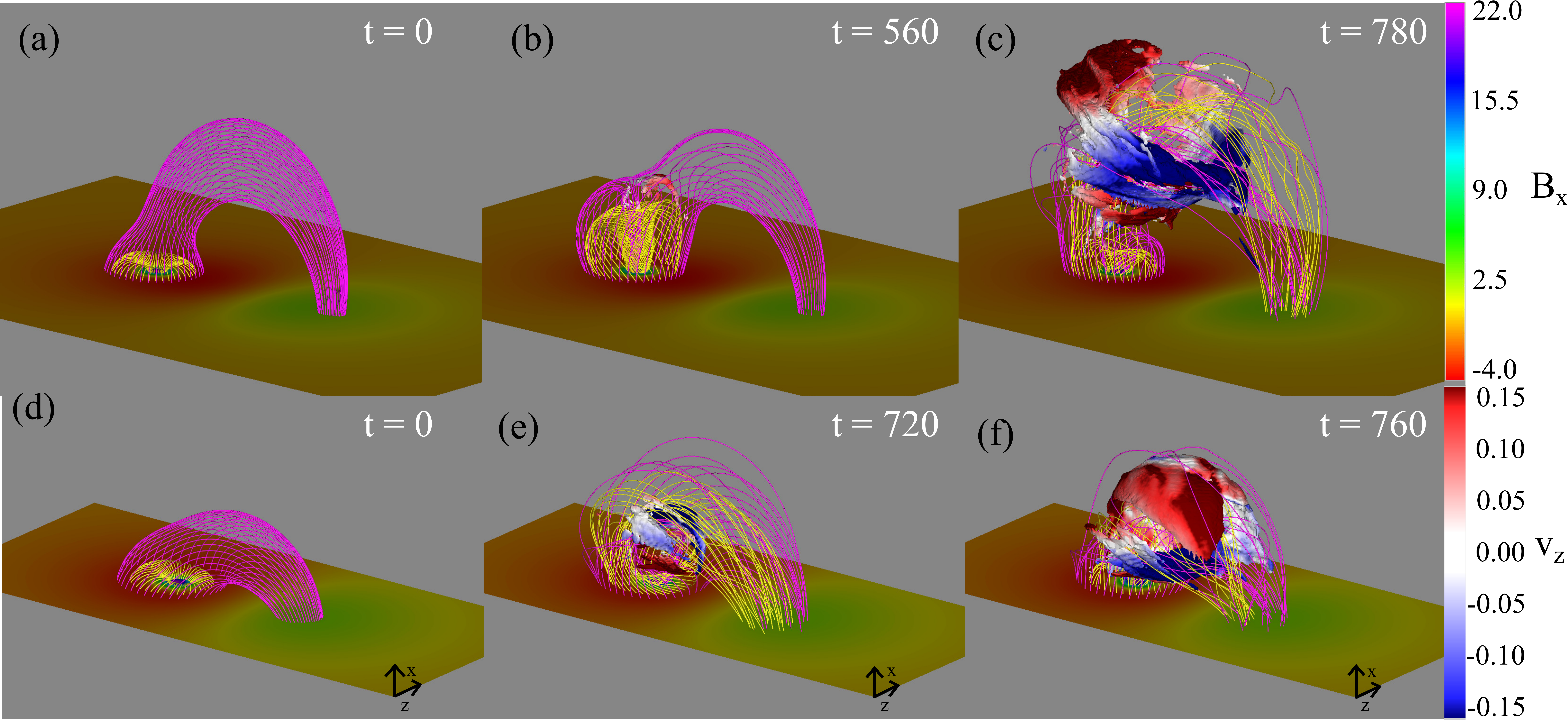}
\caption{The confined jets in each configuration. Yellow and purple field lines are traced from line-tied, non-driven footpoints just inside and outside, respectively, the separatrix surface at $t=0$. $|\mathbf{v}| = 0.3$ isosurfaces, colored according to $v_{z}$, show the rotational character of the outflows. Top panels (a-c): $L/N = 2.40$; bottom panels (d-f): $L/N = 1.46$. An animation of each jet is available online.}
\label{fig:jet}
\end{figure*}

\section{Macroscopic Evolution}
The overall evolution of our jets is qualitatively similar to the lower-resolution calculations discussed in detail in WD16. Figure \ref{fig:jet} shows how the jet outflows are guided along the connecting coronal loops. As before, in the case with $L/N=2.40$ weak outflows occur prior to the jet (Fig.\ \ref{fig:jet}(b)) which increase in intensity as the jet is launched. The jetting plasma has a {toroidal} appearance (Fig.\ \ref{fig:jet}(c)), in the manner of a subset of observed helical jets \citep[e.g.][]{Shen2011}. The jet in this case is strong and travels freely along the loop, unhindered by reflections from the far-loop footpoints. In the configuration with $L/N=1.46$, prior to the onset of the kink-like instability, relatively slow reconnection links field lines previously closed beneath the separatrix to the coronal-loop field (yellow field lines, Fig.\ \ref{fig:jet}(e)). {Although slow, this reconnection affects a significant fraction of the magnetic flux beneath the dome by the time the jet is launched.} Thereafter, the short travel time along the loop allows interactions to occur between return flows reflected along the loop from the far-loop footpoints and freshly generated jet outflows. This effect, together with the free energy drained away by the pre-jet reconnection, leads to a shorter-duration, weaker jet.

\begin{figure}[t]
\centering
\includegraphics[width=0.45\textwidth]{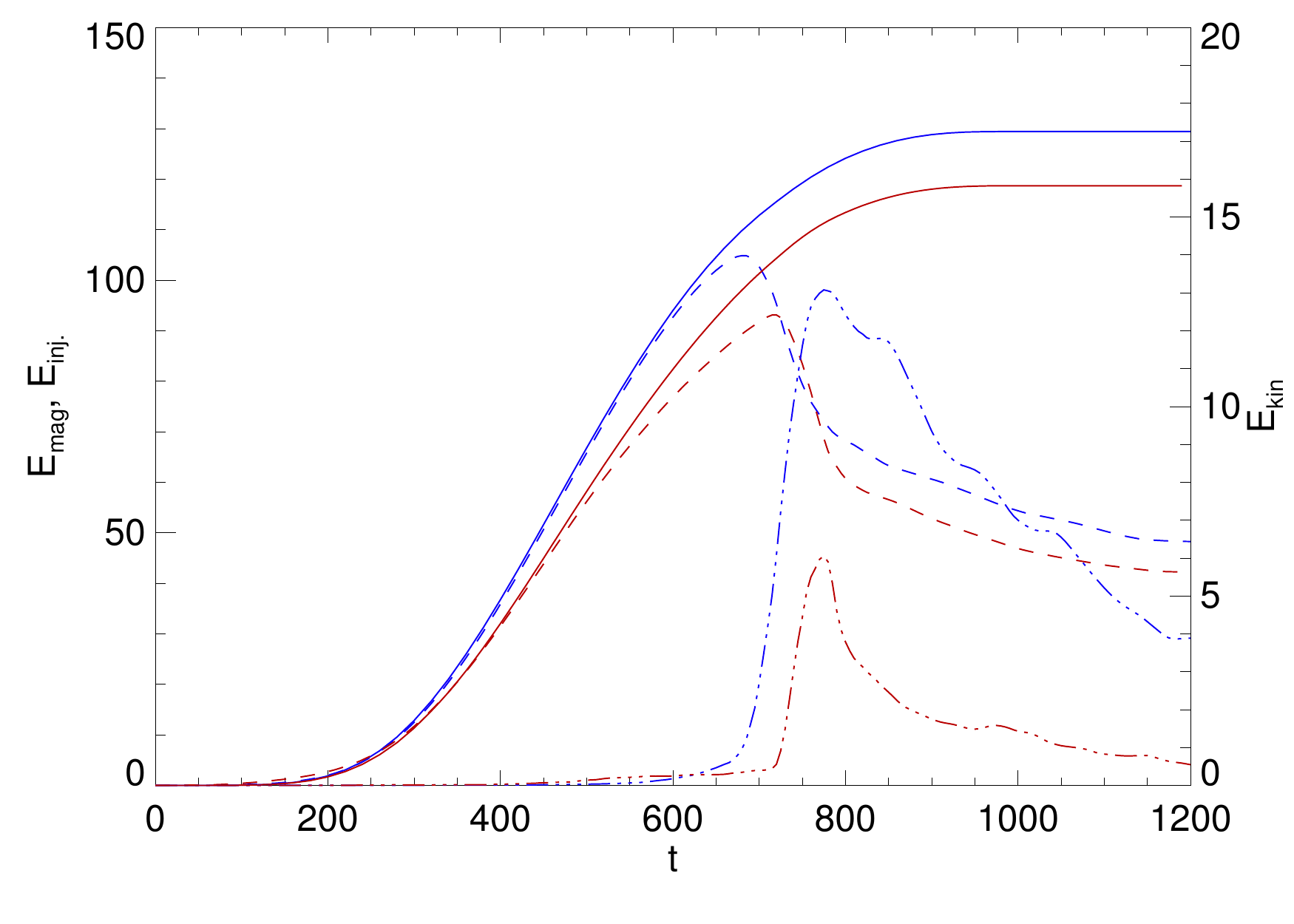}
\caption{Blue: $L/N = 2.40$, red: $L/N = 1.46$. (a) Kinetic ($E_{kin}$, triple-dot dashed) and stored magnetic ($E_{mag}$, dashed) energy evolution in each jet. Solid lines show the cumulative integrated Poynting flux ($E_{inj}$).}
\label{fig:energy}
\end{figure}

Figure \ref{fig:energy}(a) compares the volume-integrated magnetic ($E_{mag}$) and kinetic ($E_{kin}$) energies with the total energy ($E_{inj}$) injected by the boundary driving in each simulation. Both jets are marked by a sharp drop in magnetic energy, closely followed by a sharp increase in kinetic energy as the Lorentz force of newly reconnected field lines accelerates plasma near the null region. {The pre-jet reconnection plays little role in the onset of the jet for $L/N=2.40$, with fast reconnection starting explosively when the ideal kink instability is triggered. As the jet trigger depends almost entirely on this ideal process, the trigger time ($t_{trig}\approx 660$) is very similar to the lower-resolution calculation discussed in WD16. The slower reconnection prior to the jet for $L/N=1.46$ plays a more prominent role in determining when the jet is triggered, by releasing some of the stored energy (WD16). This manifests in a deviation of $E_{mag}$ from $E_{inj}$ in the energy-storage phase. Due to the better-resolved current layer and sheared-field region beneath the dome, the rate of twist accumulation beneath the dome is higher than for the lower-resolution calculation and the jet is triggered sooner ($t_{trig}\approx 720$). The peak kinetic energies of both jets are greater than in the lower-resolution calculations, consistent with the expected reduction in magnetic and viscous diffusion in the volume where the jet occurs.}

\begin{figure}[t]
\centering
\includegraphics[width=0.5\textwidth]{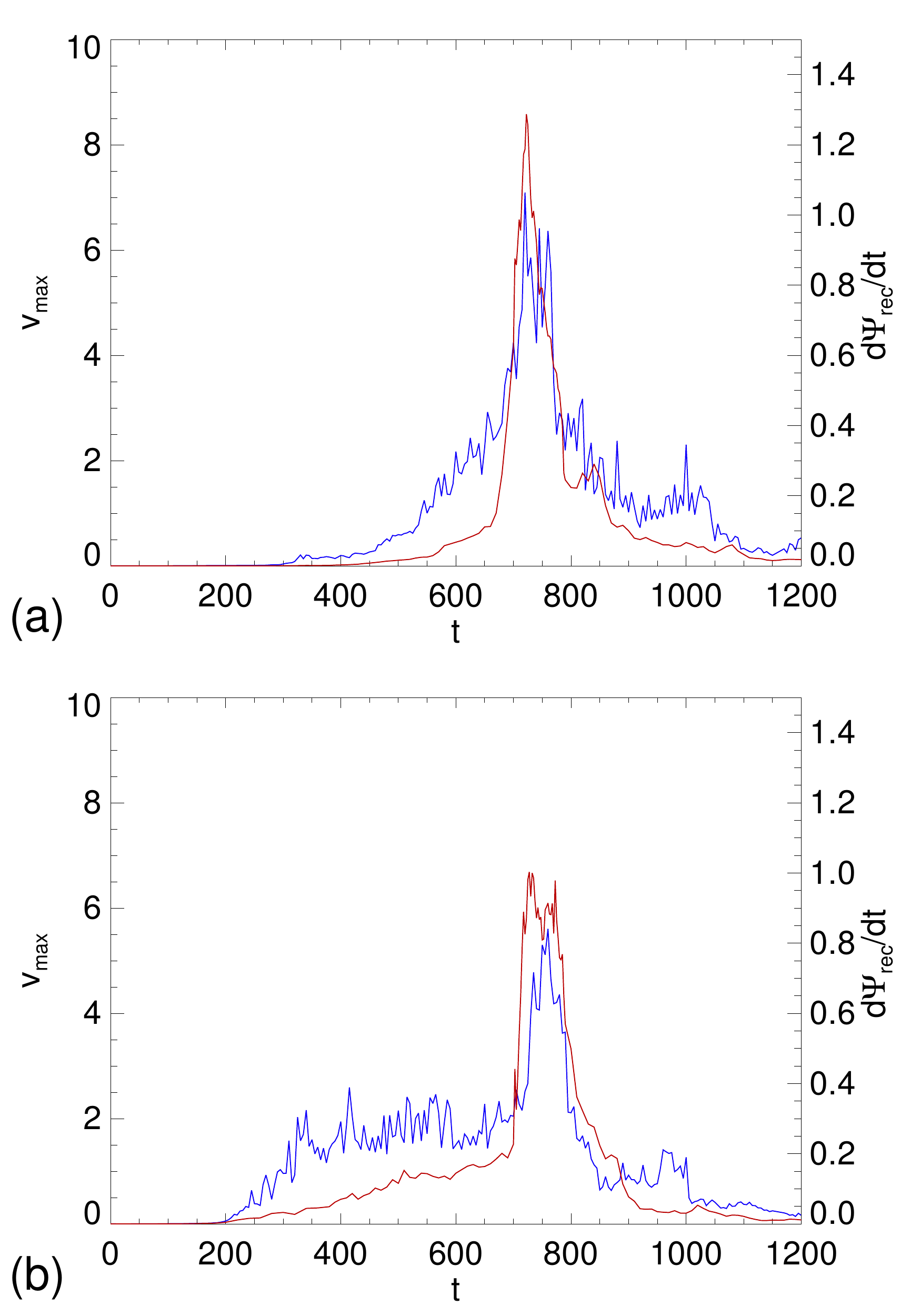}
\caption{Peak velocity magnitude $v_{\rm max}$ in the volume (blue) and the rate of interchange reconnection $d\Psi_{\rm rec}/dt$ (red) during each simulation. (a) $L/N = 2.40$, (b) $L/N=1.46$.}
\label{fig:recon}
\end{figure}

In contrast to the smooth changes in global energies, the peak velocity magnitude $v_{\rm max}$ in the volume shows much more intermittency (Fig.\ \ref{fig:recon}, blue lines), reflecting the bursty nature of the reconnection outflows. The early formation of the current layer and subsequent pre-jet reconnection for $L/N=1.46$ leads to an increase of $v_{\rm max}$ from negligible values at $t \approx 200$. The fluctuations in $v_{max}$ begin soon after this time. The fluctuating value of $v_{\rm max}$ then increases rapidly at the onset of the jet around $t \approx 720$, before decreasing once more as the jet outflows break up and interact in the connecting loop ($t \approx 800$). For the case $L/N=2.40$, $v_{\rm max}$ remains small and slowly increases as the null current layer forms and begins to reconnect. The fluctuations in $v_{\rm max}$ begin at $t \approx 550$, building to a peak value mid-way through the jet ($t \approx 720$).

For comparison, the red curves in Figure \ref{fig:recon} show the rate of interchange reconnection driven by the current layer formed around the null point. {We calculated this by tracing field lines from the photosphere, assigning each a magnetic flux element, and counting the number that cross the separatrix surface (see WD16 for details).} The two quantities have similar overall evolution, with the reconnection rate varying somewhat more smoothly. This partially results from sampling the reconnection rate in the jets at high cadence ($\Delta t = 2.5$) only around the peak times of each jet, due to the impracticality of analyzing all of the data for the entire time series. In addition, fragmented 3D reconnection is the cumulative effect of many reconnection regions \citep{Wyper2013b,Wyper2015a}, which smooths out the effect of a single burst of reconnection in the volume. Even so, the high-cadence sampling of the reconnection rate around the peaks also shows some intermittency, particularly for $L/N=1.46$.

For both jets, we conclude that the macroscopic evolution is slightly altered by the increased resolution and effective Lundquist number, but is broadly similar to the lower-resolution calculations discussed in WD16. The newly resolved tearing superimposes small-scale structure and associated intermittency on top of the macroscopic evolution of the system. This point is discussed further in $\S$ \ref{sec:discussion}. First, we investigate the fine-scale structure in more detail and discuss its possible observational signatures.

\begin{figure}[t]
\centering
\includegraphics[width=0.5\textwidth]{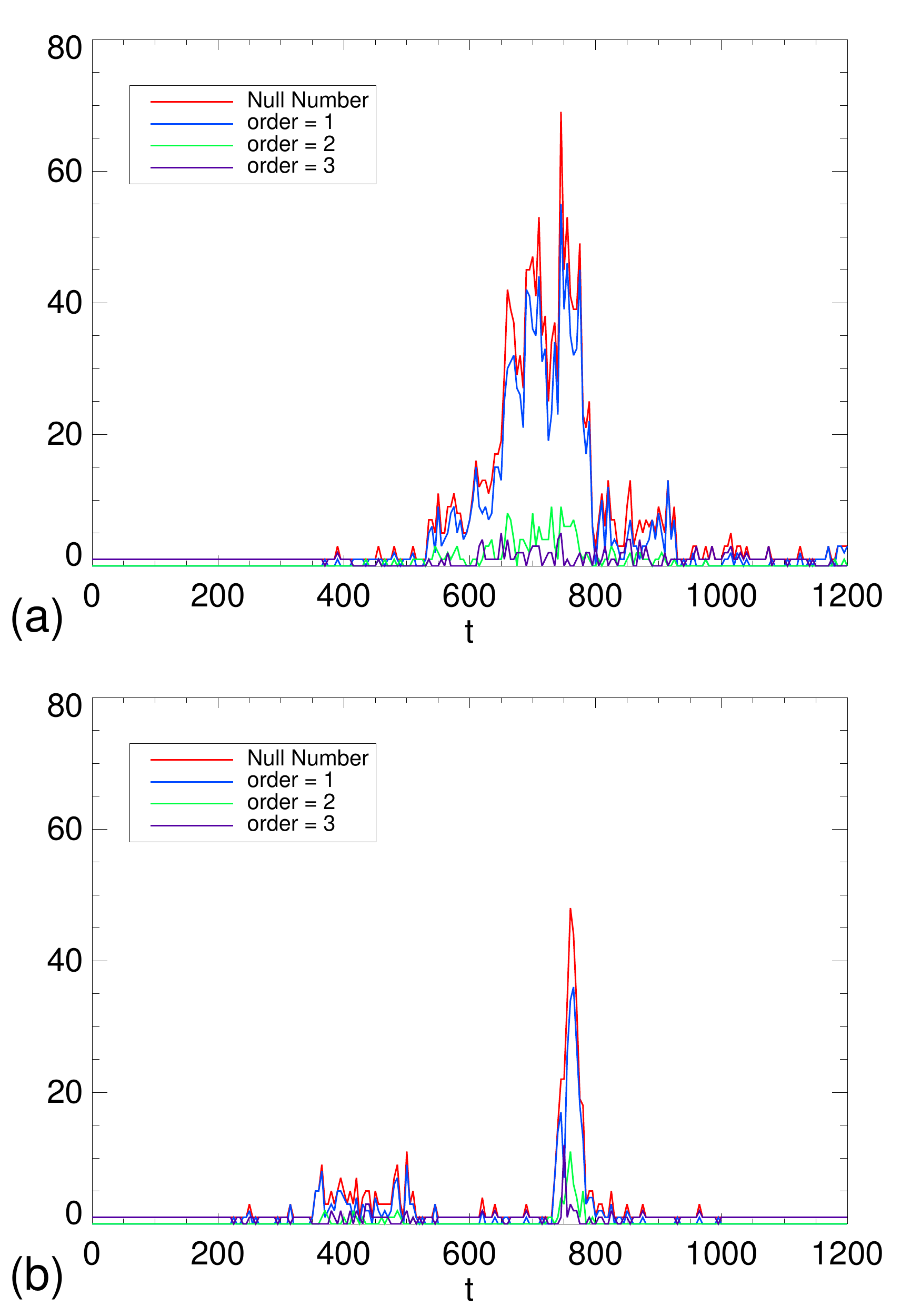}
\caption{The total number of nulls {in the vicinity of the separatrix} (red) and the number of those nulls with orders $1$ (blue), $2$ (green), and $3$ (purple), for (a) $L/N = 2.40$ and (b) $L/N = 1.46$. }
\label{fig:nullspiral}
\end{figure}
 
\section{Tearing and Intermittency}

\subsection{Nulls}
\label{sec:nulls}
The tearing-mediated reconnection in the jets occurs within a current layer formed at the pre-existing 3D coronal null point. \citet{Wyper2014a} showed how the onset of tearing in such current layers fragments the null region, forming a reconnection region that contains multiple null points and localized flux rope structures. Thus, the number and position of the null points are good indicators of the tearing that occurs during the jets.

We tracked the null points in both simulations using the trilinear method introduced by \citet{Haynes2007} (for details see Appendix \ref{ap:null}). {Nulls were identified within the current layers and also within the outflow jets. We limit our analysis to nulls formed in and around the current layer by ignoring nulls outside of a sub-volume surrounding the evolving separatrix surface in each (see Appendix \ref{ap:null}).} Figure \ref{fig:nullspiral} shows the total number of nulls, and the number of these nulls that have orders 1, 2, or 3 as a function of time. The orders represent the isolation of each null: nulls of order $1$ contain a null of opposite type in an adjacent cell, order $2$ nulls have such a null two cells away, and order $3$ at least three cells away. The formation of multiple nulls with higher orders indicates that the null region of the current layer has broken up significantly and that the null points and flux-rope structures are well resolved.

\begin{figure*}[t]
\centering
\includegraphics[width=1.0\textwidth]{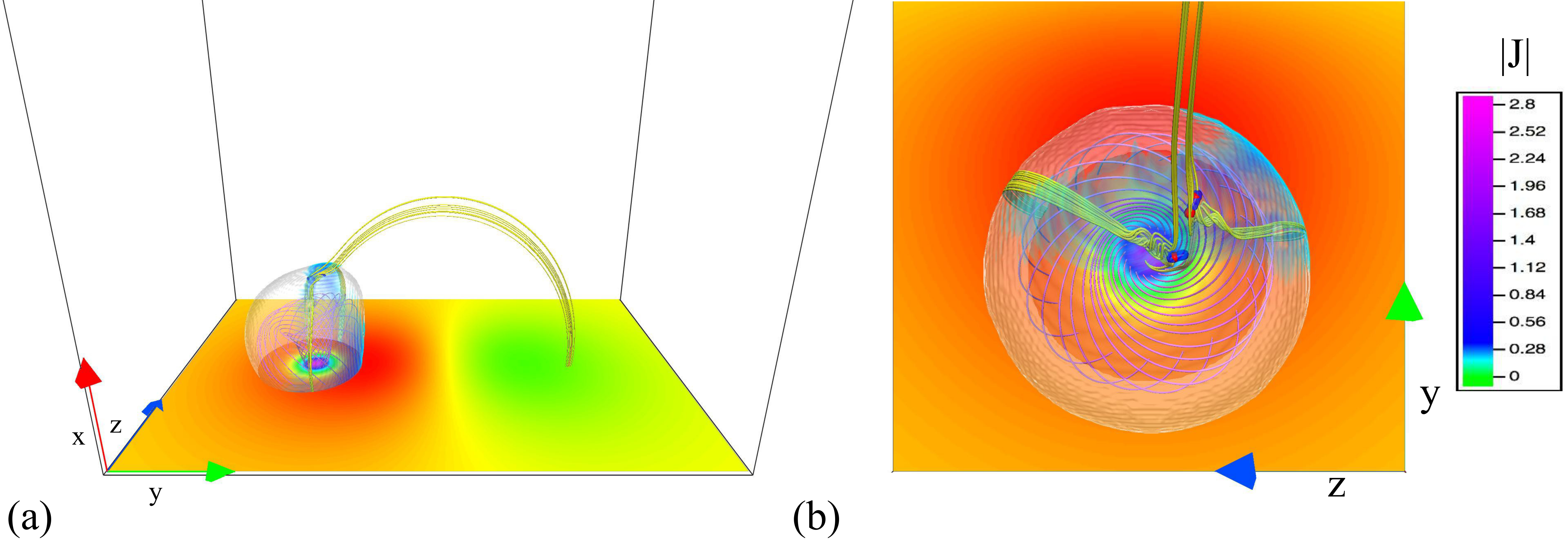}
\caption{Multiple nulls and flux ropes within the current layer at $t = 540$ {for the $L/N=2.40$ case.} (a) Side view of the field at this time. (b) Close-up view of the current layer. Semi-transparent grey isosurface shows the global separatrix between magnetic fluxes connecting to the parasitic polarity and to the far-loop footpoint. Null points shown as blue and red spheres correspond to type B and A nulls, respectively. Light blue shading shows the current density near the separatrix. Purple field lines show the twisted field beneath the separatrix dome. Yellow field lines show two flux ropes formed near the null points.}
\label{fig:1strope}
\end{figure*}

The appearance of the first nulls approximately coincides with the beginning of the fluctuations in peak velocity magnitude, as the current layer centered around the original null point begins to tear. For $L/N=2.40$, the reconnection remains focused on this region, which becomes increasingly fragmented as the jet proceeds. As an example, Figure \ref{fig:1strope} shows the positions of the nulls identified within the current layer soon after the onset of tearing. The semi-transparent grey isosurface shows the global separatrix surface, which divides the flux that connects to the parasitic polarity from the flux that connects to the far-loop footpoints (Appendix \ref{ap:separatrix} describes how this is calculated). Light blue shading shows the current layer. At this time, we identified seven nulls, which group together into two clusters residing on the separatrix surface. Shown in yellow are field lines within two nearby flux rope structures. At the most fragmented stage of the current layer evolution, over $60$ nulls were identified (Fig.\ \ref{fig:nullspiral}). The majority of these have order $1$, so they are just resolved on the grid, but up to about 10 nulls of orders $2$ and $3$ also are identified during the jet. Towards the end of the jet, {pairs of} nulls annihilate and the null region relaxes back towards a single point. 

\begin{figure*}
\centering
\includegraphics[width=0.7\textwidth]{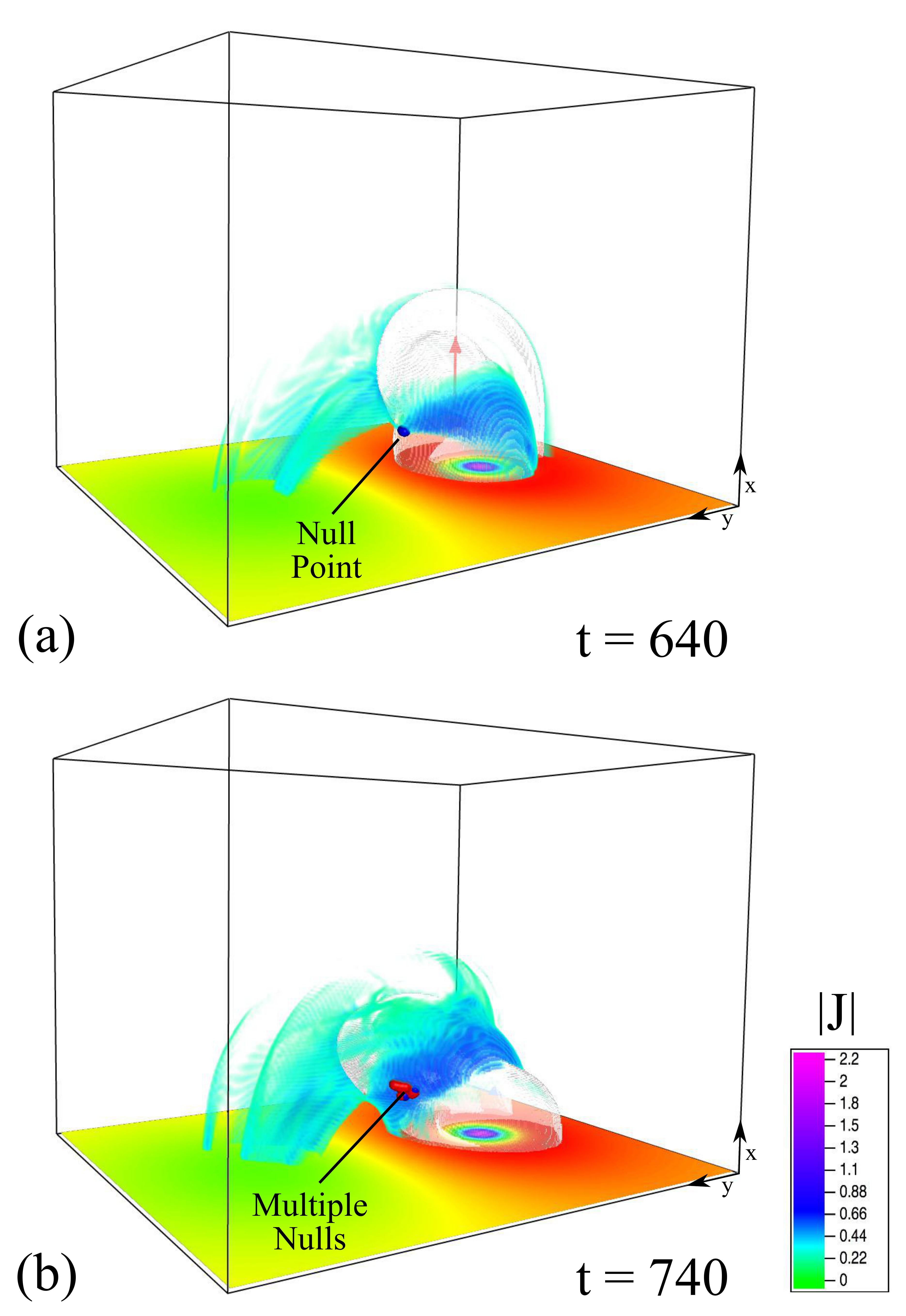}
\caption{The shift of the null position (blue sphere in (a)) back into the current layer (red and blue spheres in (b)) that initiates the fast tearing-mediated jet reconnection {in the $L/N=1.46$ case}. Shading and isosurfaces as in Figure \ref{fig:1strope}.}
\label{fig:nullshift}
\end{figure*}

The evolution of the reconnection region is rather different in the more asymmetric jet configuration with $L/N=1.46$. The current layer forms initially at the null point, where the first of the tearing begins at $t\approx 300$ (Fig.\ \ref{fig:nullspiral}(b)). This is about the same time as the fluctuations seen in $v_{\rm max}$ (Fig.\ \ref{fig:recon}(a)). However, at $t\approx 550$ the main reconnection site moves away from the null region and onto the flank of the separatrix surface. This is a fully 3D effect brought on by the highly asymmetric dome configuration, which bulges outwards directly above the parasitic polarity and {pushes the null in the direction of the far footpoint of the connecting coronal loop.} Figure \ref{fig:nullshift}(a) shows a visualization of the magnetic topology during this time. The region of high current density (light blue shading) is well away from the null (dark blue sphere) on the side of the separatrix. Also shown is some of the current present in the connecting loop following the transfer of magnetic shear by the reconnection process. The component-wise reconnection that occurs within the flank reconnection region still produces bursty outflows (Fig.\ \ref{fig:recon}(a)) and signatures of tearing (discussed below). The onset of the kink instability {at $t_{trig} \approx 660$}  causes the twisted field within the separatrix to flop over, pushing into the null region and moving the main site of reconnection back over the null (Fig.\ \ref{fig:nullshift}(b)). Once this occurs, the null region quickly fragments to form many nulls during the tearing-mediated jet-reconnection phase (Fig.\ \ref{fig:nullspiral}(b)). Following the jet, the nulls coalesce and the system relaxes towards a configuration containing one null point.

\begin{figure}
\centering
\includegraphics[width=0.4\textwidth]{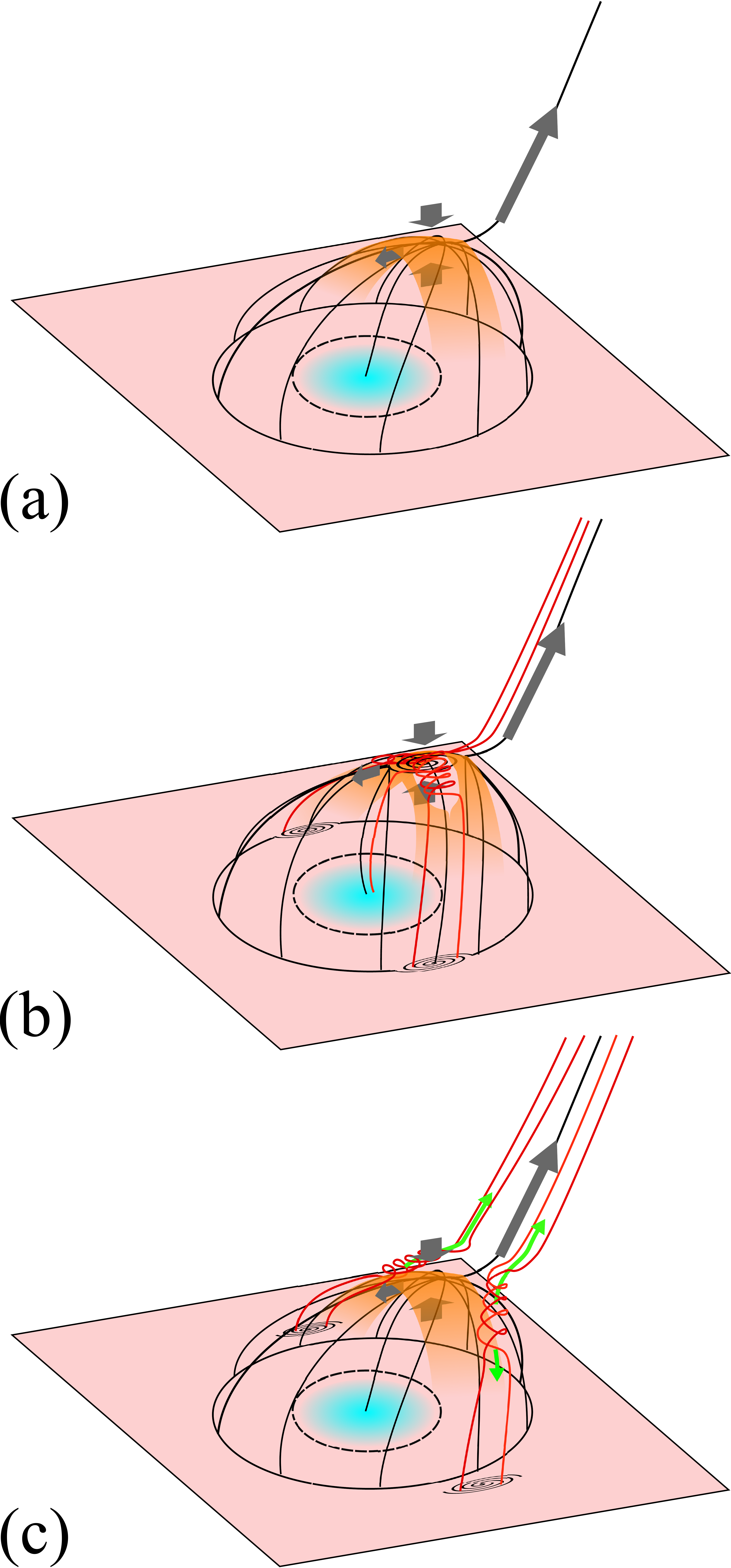}
\caption{{Schematic of tearing-mediated jet evolution, showing the reconnecting current layer (orange), quasi-steady inflows and outflows (grey arrows), twisted flux ropes in the layer and threads in the outflow (red), and propagating torsional Alfv\'{e}n waves (green). See text for details.}}
\label{fig:cartoon}
\end{figure}

\subsection{Flux Ropes}
\label{sec:ropes}
We have demonstrated that the evolution of the null points helps to identify when tearing occurs in the current layer and where the main reconnection region is located. {However,} direct observable evidence of tearing tends to focus on the formation of blob-like structures assumed to be associated with magnetic islands \citep[e.g.][]{Zhang2014}. The 3D equivalent of islands -- flux ropes -- form repeatedly in our jets. Here, we explore in detail one example of flux-rope formation and evolution, and show how the formation of flux ropes can be diagnosed using quantities mapped on the photosphere. We also show that their evolution leads to a surprisingly complex final state.

{A simplified view of the flux-rope formation and evolution process is given in Figure \ref{fig:cartoon}, a schematic diagram of our tearing-mediated jet evolution. The reconnecting current layer (orange) produces quasi-steady flows (grey arrows) directed toward the layer from inside and outside of the dome, and away from the layer along the spine lines of the null (Fig.\ \ref{fig:cartoon}(a)). Increasing stress on the null, due to the evolution of the magnetic field below, lengthens the current layer until it reaches the threshold for the plasmoid instability. At this point, flux ropes (red) begin to form in the current layer (Fig.\ \ref{fig:cartoon}(b)). As the flux ropes are ejected, their twist propagates as torsional Alfv\'{e}n waves (green arrows), and twisted filamentary threads (red) within the jet outflow are created (Fig.\ \ref{fig:cartoon}(c)). The evacuation of the flux ropes from the current layer allows new ropes to form, grow, and depart, as steps (b) and (c) of the sequence repeat.}

\begin{figure*}
\centering
\includegraphics[width=0.9\textwidth]{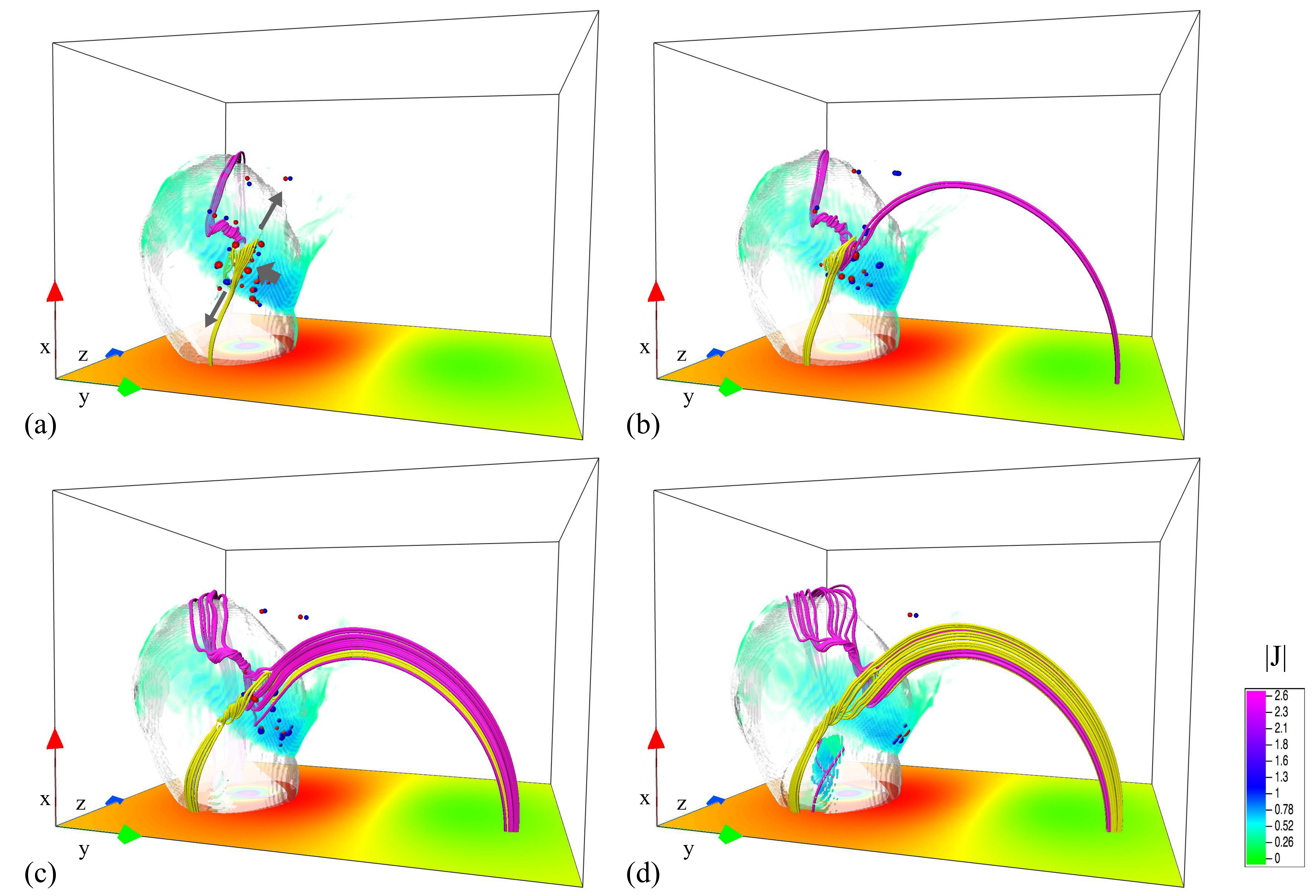}
\caption{Formation and ejection of two flux ropes during the $L/N=2.40$ jet. (a-d) show $t = 707.5, 712.5, 717.5$ and $722.5$, respectively. Representative field lines that show the rope structure at each time, but do not maintain their identity between frames, are plotted in {magenta} and yellow. {Grey arrows in (a) show the direction of plasma inflow and outflow from the current layer.} Shading, isosurfaces, and null points as in Figure \ref{fig:1strope}.}
\label{fig:expel}
\end{figure*}

Figure \ref{fig:expel} shows the evolution of two flux-rope structures in the current layer of our $L/N=2.40$ simulated jet. {The current layer is localised to the vicinity of the separatrix surface, wrapping around it so that the outflows are angled nearly vertically upwards and downward towards the photosphere around this time (Fig.\ \ref{fig:expel}(a)).} During the jet, the current layer rotates around the parasitic polarity, as sheared field inside the separatrix sequentially reconnects with unsheared field of the coronal loop outside the separatrix. The two flux ropes form initially in the centre of the fragmented null region (Fig.\ \ref{fig:expel}(a)), where the twist within each is concentrated into a tight bundle. At this time, the layer is already broken up, and any symmetry perpendicular to the outflow direction in the layer has been lost. The ends of the field lines within each rope initially connect to flux within the separatrix, i.e. the flux ropes form on the underside of the current layer wrapping around the separatrix surface. As the jet proceeds, one end of each flux rope {locally} opens up by reconnecting with the coronal-loop field (Figs.\ \ref{fig:expel}(b)-(c)). Also around this time, the twist within each rope begins to relax and spread out along the length of the field lines, whilst the ropes start to wrap into one another. This braiding and twist propagation is fully consistent with the results of \citet{Wyper2014b}. As the reconnection region moves on to process more flux, the two ropes are left behind as twisted threads within the coronal loop. The twist within each thread then begins to distribute itself evenly along the length of the thread (Fig.\ \ref{fig:expel}(d)). Thus, for our helical jets, which are themselves driven by a large-scale torsional wave pulse, tearing leads to small-scale torsional wave packets that propagate outwards as part of the jet curtain.

{This concentration and relaxation of twist appears to be a universal process that occurs repeatedly throughout our jets. Our localized flux ropes have between $1$ and $5$ turns, depending upon where they form and their lifetimes within the layer. The lifetimes range from $\approx 12.5$ to $\approx 25$ time units for the largest flux ropes, which are ejected over an interval comparable to their lifetime. The fastest travel near the inflow Alfv\'{e}n speed ($\approx 0.3$) as they exit the current layer. Note that at increased resolution, we would expect smaller, shorter-lived flux ropes to form between these larger ones. As discussed in \citet{Wyper2014b} there is no one-to-one correspondence between the flux ropes and the null points in the layer when it is fragmented. In our case, there are many more nulls than significantly large flux ropes, with the nulls congregating around the ropes when they form in the weak-field region at the center of the current sheet (e.g., Fig.\ \ref{fig:expel}(a)).}

Tightly wound, thin flux ropes such as those described above involve only a small amount of magnetic flux. However, much larger flux ropes that involve much more magnetic flux also form, and spread over a large extent of the current layer that wraps the separatrix surface. These flux ropes are typically not tightly wound and so are difficult to identify based on plotting field lines. However, they contain sufficient flux to be readily identified by examining the connectivity of the magnetic field. Specifically, they appear as swirls or spirals in the squashing factor $Q$ \citep{Titov2002,Titov2007} evaluated on the photosphere {(Appendix \ref{ap:q} describes the procedure used to calculate Q in our simulations).} The squashing factor highlights gradients in the magnetic field mapping from each point on the photosphere. Q is formally infinite at the footpoints of separatrix surfaces and spine lines, and is very large at the footpoints of field lines that trace into quasi-separatrix layers (QSLs). As the largest flux ropes form, they locally twist up field lines to produce spirals in $Q$, particularly when this local twisting cuts across a layer of high $Q$ such as a separatrix surface \citep[e.g.][]{Pontin2015} or QSL associated with filamentary current layers in the loop. 

\begin{figure*}[t]
\centering
\includegraphics[width=0.95\textwidth]{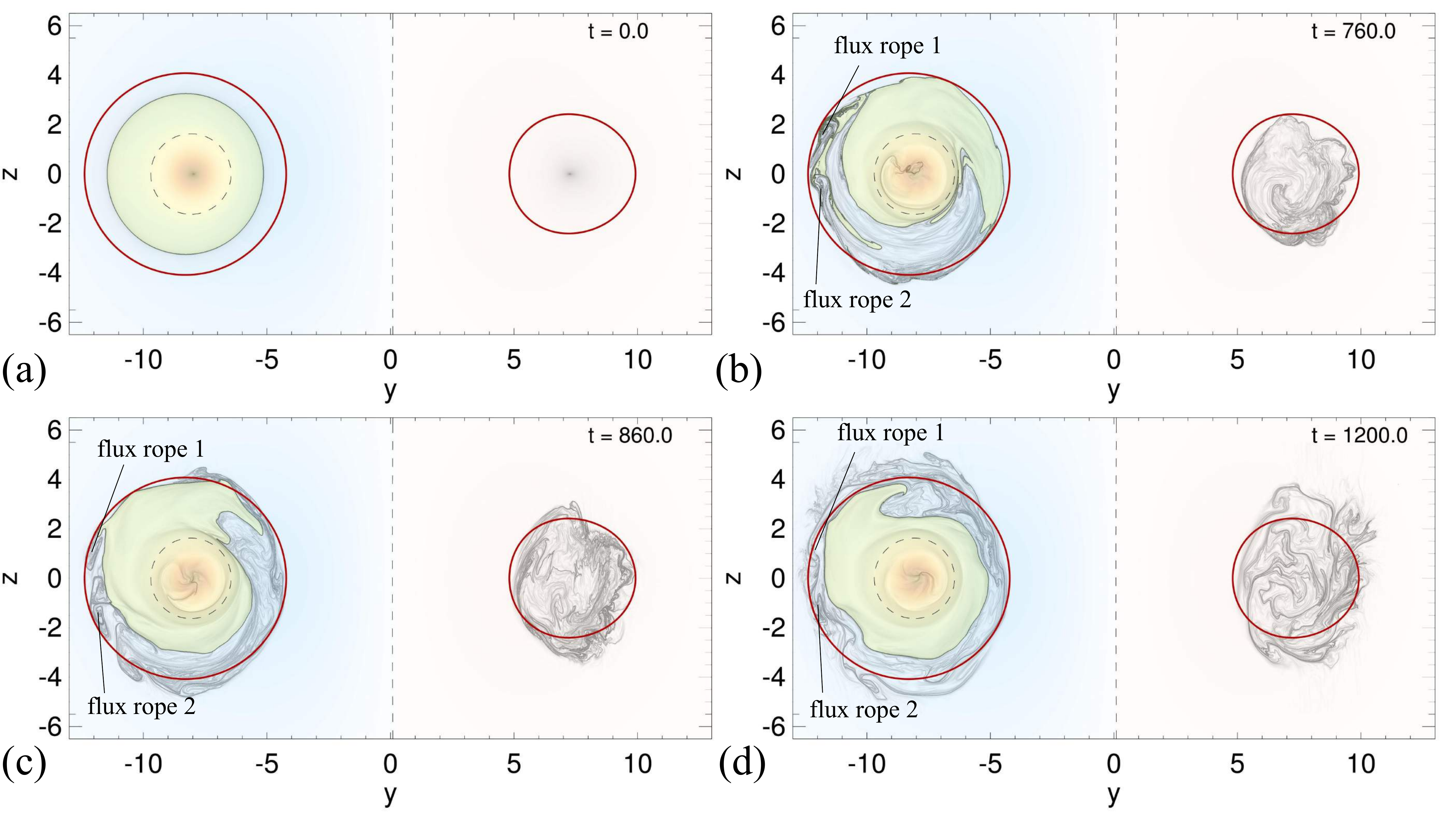}
\caption{{Evolution of the squashing factor ($Q$, greyscale) on the photosphere during the $L/N=2.40$ jet. Also shown are the polarity inversion lines (dashed lines) and the magnetic field component normal to the photosphere (red and blue indicate positive and negative polarity, respectively). Green shading shows the flux beneath the separatrix dome. Thick red lines show the footprint of the flux tube within which the jet is predicted to be contained (see text for details). An animation of this figure is available online.}}
\label{fig:qbp8}
\end{figure*}

Figure \ref{fig:qbp8} shows the formation and relaxation of two large flux ropes, which we denote flux ropes $1$ and $2$, in the $L/N=2.40$ jet. $Q$ is shown in grey scale, whilst dashed lines depict the polarity inversion lines. Red and blue contour shading shows the normal component of the magnetic field at the photosphere. To differentiate between the two flux systems, flux that connects to the parasitic polarity and resides within the separatrix is shaded green. Both flux ropes form around the time of the peak jetting, when the current layer is most extended. 

Flux rope 1 forms on the underside of the current layer in the locally closed field. {The spiral shape of the open/closed boundary shown in Figure \ref{fig:qbp8}(b) is formed as this large flux rope is opened up by the global interchange reconnection occurring during the jet, in the same manner as the two smaller flux ropes discussed above. In contrast, flux rope 2 spans the separatrix at formation and creates the spiral shape in the open/closed boundary at the photosphere directly, in the manner assumed by the static model studied by \citet{Pontin2015} (see also the online animation). As the global interchange reconnection continues,} both flux ropes are soon entirely reconnected with the coronal-loop field, where they appear as flattened spirals in $Q$ and form twisted threads within the loop (Fig.\ \ref{fig:qbp8}(c)). By the end of the simulation, some of the twist within each thread has been redistributed by reconnection within the loop, and the spirals have partly unwound (Fig.\ \ref{fig:qbp8}(d)).

\begin{figure*}[t]
\centering
\includegraphics[width=0.95\textwidth]{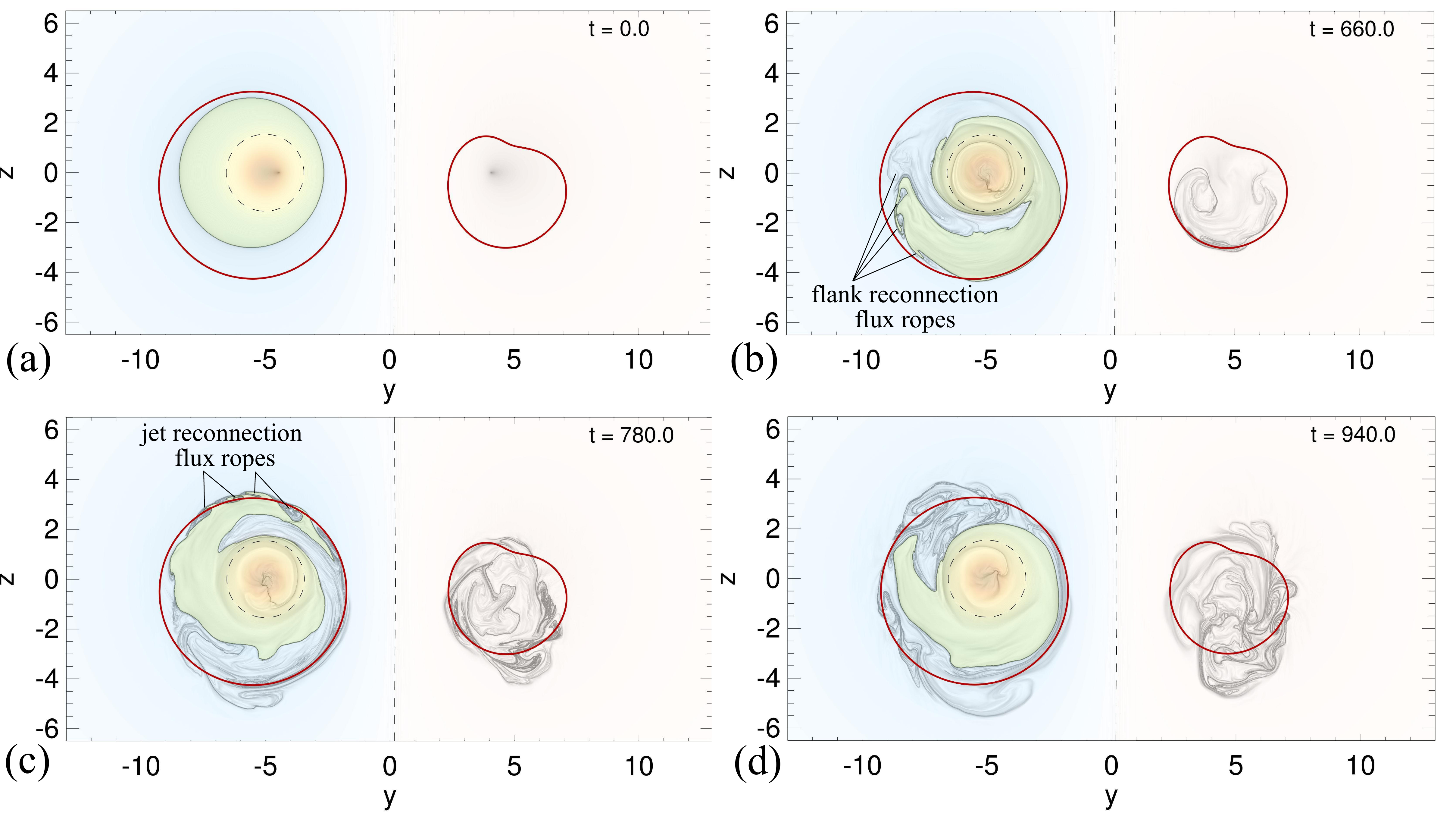}
\caption{Evolution of the squashing factor ($Q$, greyscale) on the photosphere during the $L/N=1.46$ jet, with color scales and the polarity inversion lines shown as in Figure \ref{fig:qbp8}. An animation of this figure is available online.}
\label{fig:qbp5}
\end{figure*}

In the $L/N=1.46$ jet, similar large-scale flux ropes are evident and wrap over a large extent of the separatrix surface. As discussed in \S \ref{sec:nulls}, the reconnection site moves away from the null onto the flanks of the separatrix surface prior to the jet. Figure \ref{fig:qbp5}(b) shows that during this time ($550 \lesssim t \lesssim 730$), multiple large flux ropes form, spanning the separatrix surface and sequentially reconnecting onto the coronal-loop field (see also the online movie). The formation of these flux ropes shows that tearing is indeed occurring during this time and helps to explain the bursty nature of $v_{\rm max}$ measured in the volume (Fig.\ \ref{fig:recon}(a)) throughout this interval. A similar sequence of large-scale flux ropes is produced later during the jet phase (Fig.\ \ref{fig:qbp5}(c)), and their spiral $Q$ layers also then smooth out and unwind as the field relaxes (Fig.\ \ref{fig:qbp5}(d)).

\begin{figure}
\centering
\includegraphics[width=0.45\textwidth]{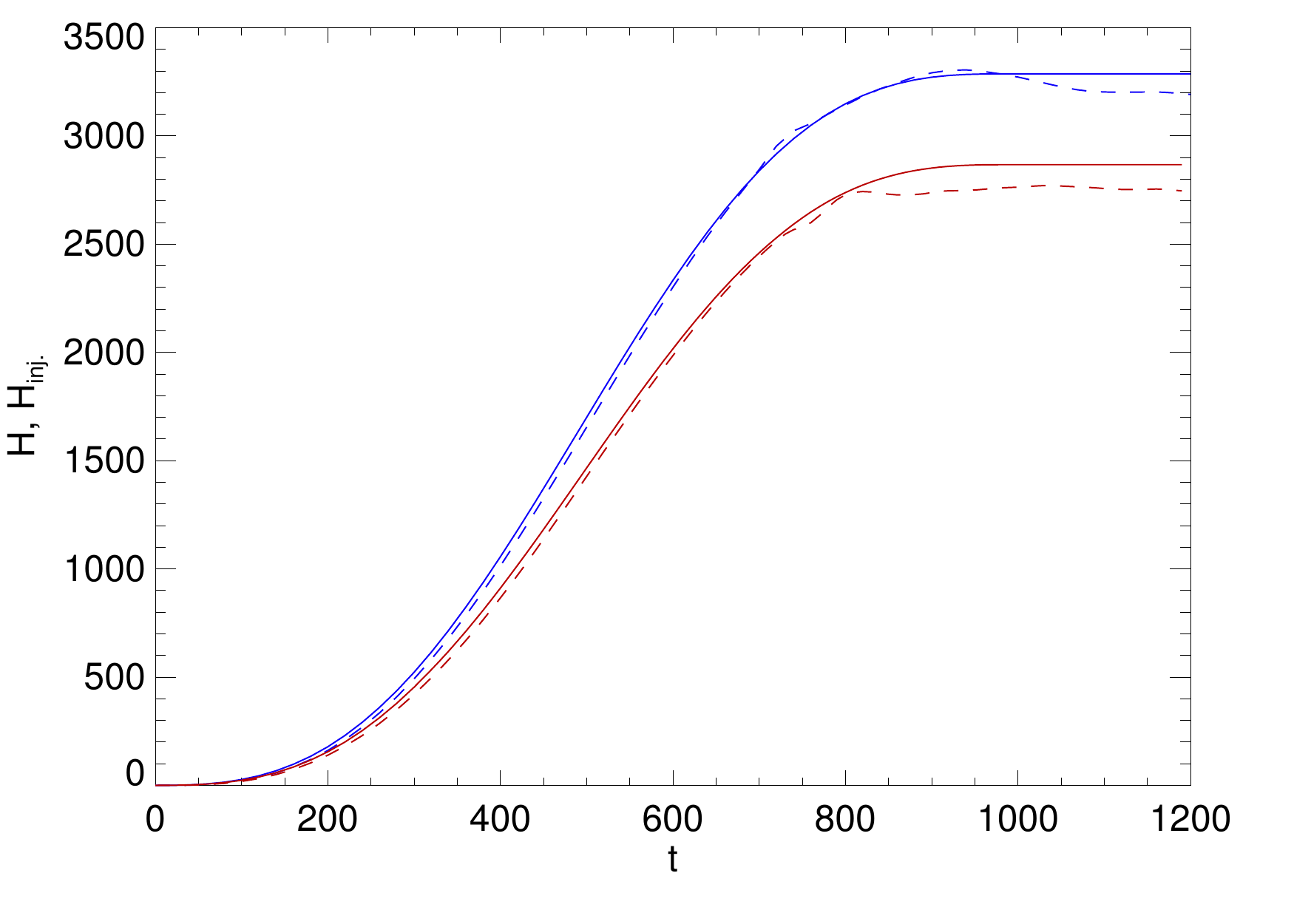}
\caption{The injected (solid) and volume integrated (dashed) helicity in each jet. Blue: $L/N=2.40$, red: $L/N=1.46$.}
\label{fig:helicity}
\end{figure}

The filamentary structure of $Q$ also shows the extent and position on the photosphere of the magnetic flux affected by the jet. This structure should correspond to areas of enhanced brightening, in response to energy deposition by heat flux and high-energy particles associated with the jet reconnection. Over the duration of the energy-buildup and -release phases of both jets, the flux of the parasitic polarity is reconnected roughly twice: once when the twisted field is reconnected onto the coronal loop (where it imparts some of its twist and helicity), and again when the field is reconnected back down to close beneath the separatrix surface. This process brings the whole configuration closer to a minimum-energy state consistent with the helicity injected into the system. {Since the second reconnection phase is a repeated reconnection of flux that was previously beneath the separatrix dome} (for a detailed description see WD16), the amount of affected flux {within the coronal loop} is roughly equal to the flux of the parasitic polarity. In that case, the filamentary layers of $Q$ on the photosphere should form within a flux tube that is centered on the footprint of the global separatrix surface and contains the same amount of magnetic flux as the parasitic polarity.

Figure \ref{fig:qbp8} shows the footprint of such a flux tube for the $L/N=2.40$ jet {as thick red lines (where the field lines are traced at $t=0$).} Until the late stages of the jet evolution, the reconnection and inferred associated photospheric brightening are contained within this flux tube (Fig.\ \ref{fig:qbp8}(a)-(c)). At later times, as multiple reconnection events begin to relax the filamentary structure within the coronal loop, the affected magnetic flux extends outside of this volume into the surrounding field (Fig.\ \ref{fig:qbp8}(d)). A similar evolution occurs in the $L/N=1.46$ jet. In this case, turbulent reconnection and flow interaction begin even whilst the jet is launched. To fit the ensuing $Q$ map better, we shifted the circular footprint of the flux tube to lie along the edge of the separatrix surface just prior to the jet (Figs.\ \ref{fig:qbp5}(a)-(b)). During the early phases of the jet, the affected flux is contained reasonably well within the flux tube (Fig.\ \ref{fig:qbp5}(c)), whilst later on the affected region broadens as the coronal-loop field relaxes (Fig.\ \ref{fig:qbp5}(d)).

\begin{figure*}
\centering
\includegraphics[width=1.0\textwidth]{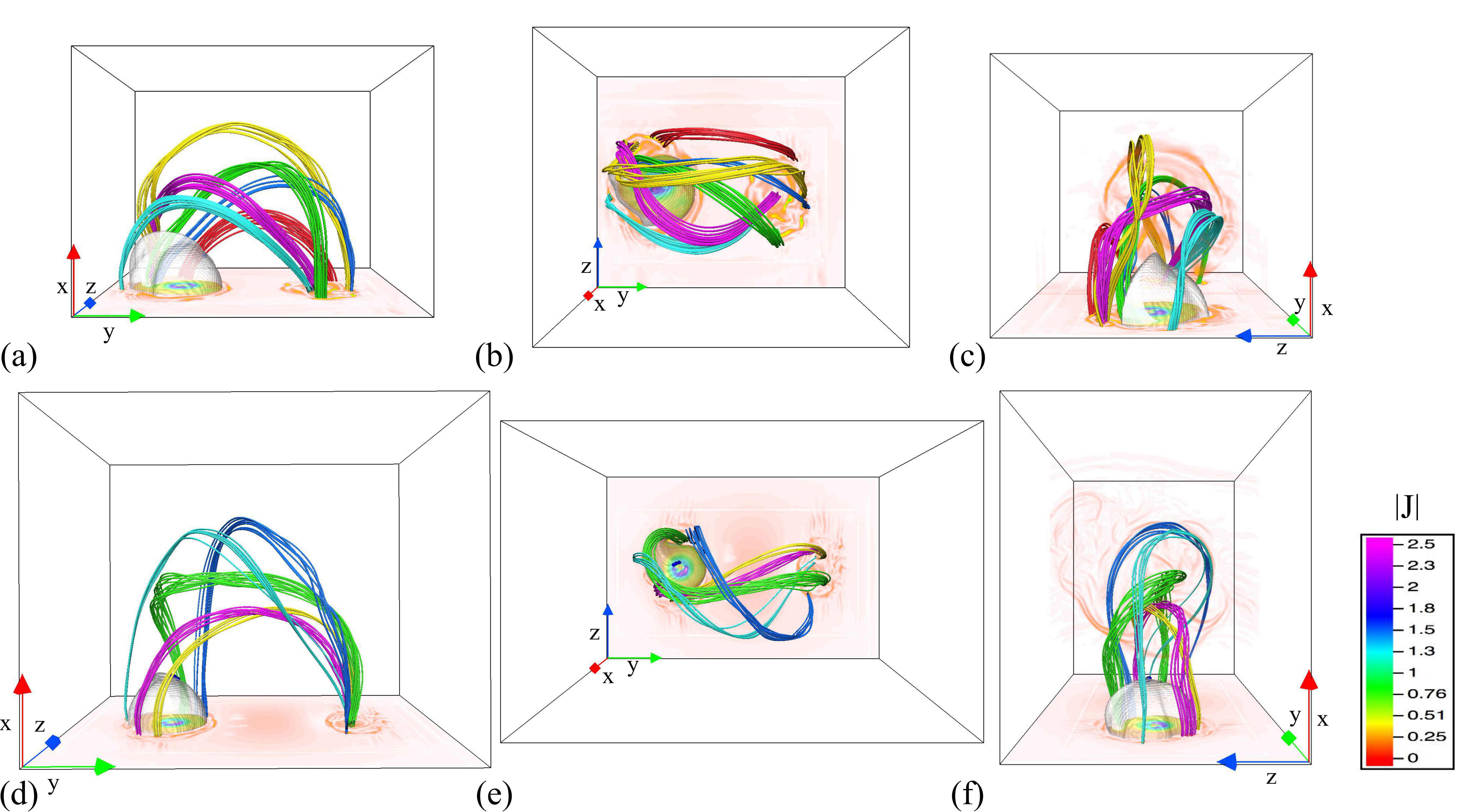}
\caption{Twisted thread-like flux ropes in the post-jet fields of both configurations {viewed from the side (left), above (middle), and along the loop (right).} (a-c) $L/N=1.46$ at $t = 940$, (d-f) $L/N=2.40$ at $t = 1200$. Semi-transparent grey isosurfaces show the separatrix surfaces. Shading on the photosphere {in each panel and} in the cross section of the coronal loop in (c) and (f) corresponds to current density magnitude.}
\label{fig:ropeend}
\end{figure*}

This formation of fine-scale structure in the field around the jetting region also must occur in coronal-hole jets. In that case, the twist and helicity within each thread can simply propagate outwards into the heliosphere. In our closed-field jets, the twist and helicity are trapped within the connecting coronal loop. Since helicity is approximately conserved during 3D reconnection \citep[e.g.][]{Priest2014}, it should be approximately conserved during the jet-generation process. Figure \ref{fig:helicity} shows that this is indeed the case for both our jets (see Appendix \ref{ap:helicity} for details of the helicity calculation). Figure \ref{fig:ropeend} shows selected field lines from the relaxed state following the jets. The remnants of tearing-generated flux ropes are seen as twisted threads that wrap around one another within the large-scale twist transferred to the coronal loop. Each thread is separated from the next by a current layer extending along the loop length, shown in cross section in Figure \ref{fig:ropeend}(c) and (f). Somewhat surprisingly, particularly given the turbulent evolution that occurs for $L/N=1.46$, neither final state takes the form of a uniformly twisted force-free loop. This suggests that although multiple reconnection episodes occur in the loop as the system relaxes, the evolution is not sufficiently turbulent or volume-filling that the \citet{Taylor1986} relaxation theory applies. With its multiple current layers separating the different twisted threads within the loop, the final state in both configurations resembles much more a 3D version of the reconnection-driven current filamentation discovered in 2.5D by \citet{Karpen1996}. In that process, tearing leads to a local misalignment of flux surfaces and the corresponding formation of long-lived currents within the coronal loop.

\begin{figure*}
\centering
\includegraphics[width=0.9\textwidth]{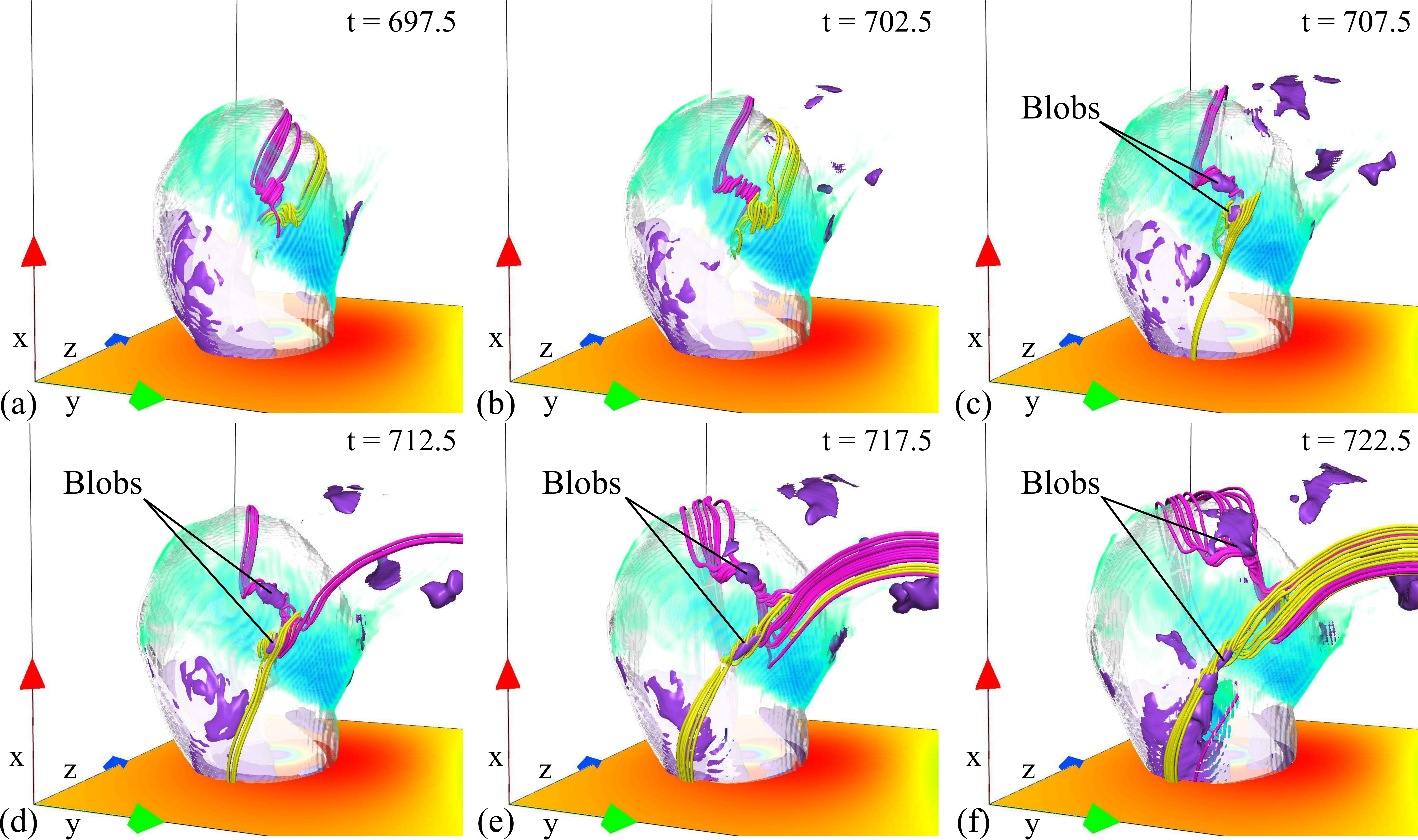}
\caption{Plasma blobs in the current layer during flux-rope formation in the weak-field jet region. {Field lines, photospheric shading and volume rendering of the current density and separatrix surface are the same as Figure \ref{fig:expel}.} Shown in purple are isosurfaces of $\rho = 1.5$ depicting plasma blobs. An animation of this figure is available online.}
\label{fig:roperho}
\end{figure*}

\subsection{Observational Signatures}
We now consider additional possible observational signatures of the tearing-generated fine-scale structure described above. As previously mentioned, the ejection of blob-like features has been interpreted as a signature for tearing-mediated reconnection in observed jets \citep[e.g.][]{Zhang2014}. Two-dimensional jet simulations have shown that large, dense islands can form and be ejected following the fragmentation of the jet current layer \citep[e.g.][]{Karpen1995,Yokoyama1996}. We have shown that in 3D such islands become flattened flux-rope structures that form both in the weak-field regions involving null points and along the flanks of the separatrix surface. Intuitively, one might expect that, in the weak-field region at least, the tightly wound flux ropes would behave in a similar way to the 2D scenario.

Figure \ref{fig:roperho} shows an isosurface of plasma density for the two flux ropes in the null region shown in Figure \ref{fig:expel}. {The isosurface is at a value of $1.5$, $50\%$ above the background. As the flux ropes form, they contract enough that regions of enhanced density (``blobs'') develop on each tightly wound rope. The growing enhancement becomes visible as blobs in the density isosurface in Fig.\ \ref{fig:roperho}(c).} However, as the twist in the ropes spreads out along the length of their field lines, so does the region of enhanced density (Figs.\ \ref{fig:roperho}(d)-(e)). The blobs are then assimilated into the higher-density regions of the current-sheet outflow (Fig.\ \ref{fig:roperho}(f)). It is less clear whether any density enhancement occurs in the larger flux ropes that form across the flanks of the separatrix surface. Their extended length and contorted shape made it difficult to identify them in the volume and assess the density variation in their vicinity as they evolved. Nevertheless, for the weak-field region at least, we have shown that blobs of enhanced density form within the current layer before becoming part of the main jet outflow. The twisted threads formed by these flux ropes within the loop have densities enhanced above that of the background. As optically thin emission is proportional to the square of the plasma density, the emission in the jet outflow then should exhibit a filamentary structure due to these density features.

{We estimate some expected observed properties of these blobs by adopting typical values for the length scale, field strength, and plasma density observed in solar jets (see WD16 for details). Choosing values of $B_{s} = 10 \,\text{G}$, $\rho_{s} = 10^{-14} \text{g cm}^{-3}$, and $L_{s} = 10^{8} \,\text{cm}$ gives jets typical of those observed in active regions. The corresponding lifetimes of the blobs then range from $12.5 - 25 \,\text{s}$ during a jet lasting $180$ s or so (when $L/N=2.40$). The fastest blobs exit the current layer at $\approx 300 \,\text{km s}^{-1}$ with these scalings. Their speeds are consistent with typically observed blob speeds, $\approx 120 - 450 \,\text{km s}^{-1}$, while their lifetimes are somewhat shorter than those reported, $\approx 24 - 60\, \text{s}$ \citep{Zhang2016}. This is consistent with the fact that the observed values include the interval whilst the blobs are still visible within the jet curtain/spire. Densities and temperatures are enhanced above the uniform background ($\rho = 10^{-14}  \text{g cm}^{-3}$, $\text{T} = 1 \text{MK}$) by a factor of $\approx 2$ and $\approx 1.5$, respectively. However, a proper treatment of the plasma energetics is required to obtain accurate enhancements and to compare to observations. Typical field strengths within the flux ropes are $\approx 0.8$, or $\approx 8 \,\text{G}$ with these scalings. Shorter lifetimes and lower typical speeds have been reported in 2.5D simulations \citep{Yang2013,Ni2015} and observations \citep{Singh2011} of chromospheric jets. By their nature, these jets tend to be smaller in size and occur in plasma where the characteristic speeds are lower. The 2.5D calculation by \citet{Yang2013} included a more realistic atmosphere and a full energy equation. Hot, dense, plasma blobs trapped within magnetic islands formed periodically in the layer and were ejected at about the local Alfv\'{e}n speed ($\approx 30 \,\text{km s}^{-1}$). Our calculation shows how this picture is altered in 3D, where the islands are localized flux ropes that form untwisting dense threads within the jet outflow, and in the corona, where the characteristic flow speeds are about an order of magnitude higher.}

\begin{figure*}[t]
\centering
\includegraphics[width=0.99\textwidth]{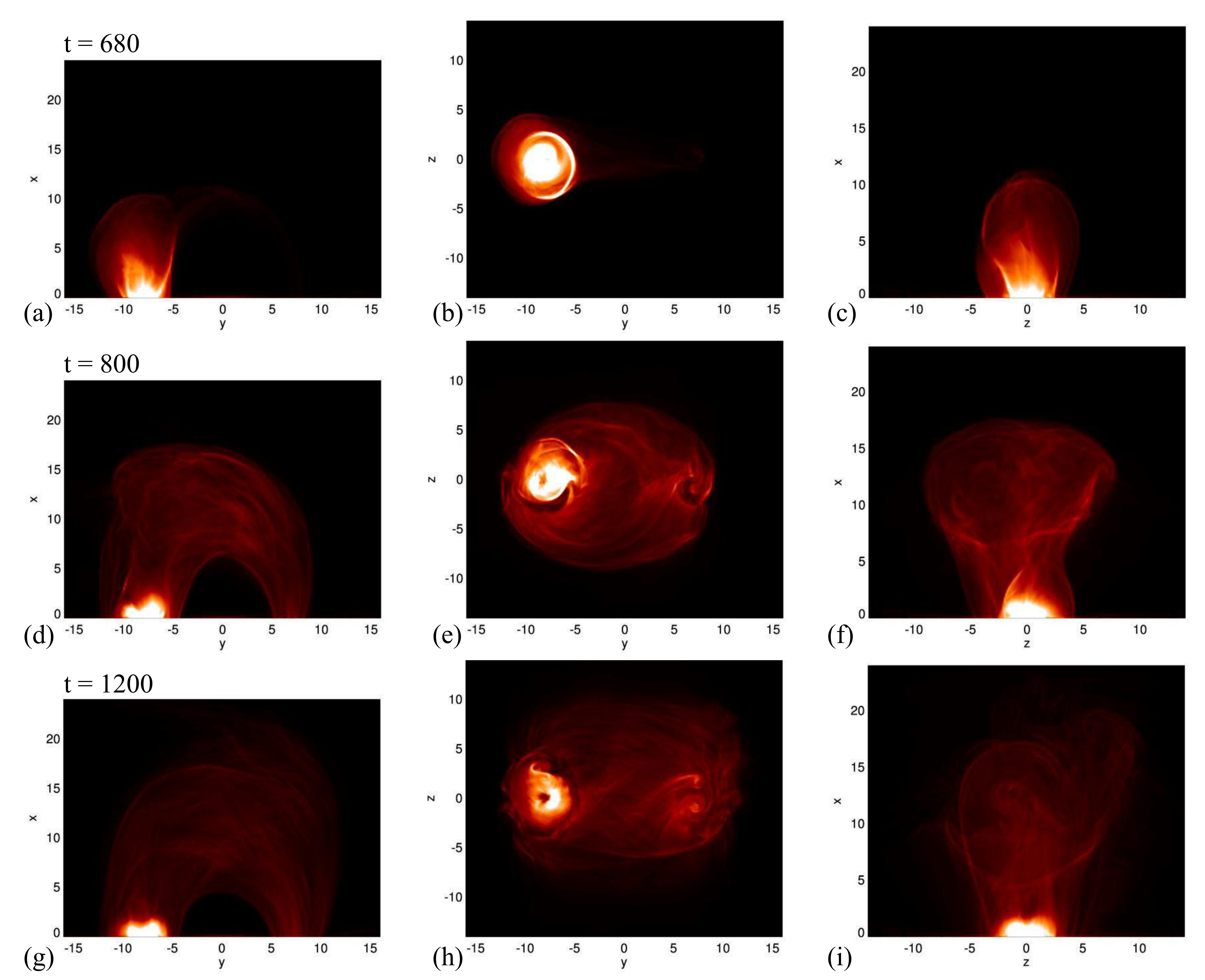}
\caption{{$L/N=2.40$. Line-of-sight integrated current density before ($t = 680$), during ($t = 800$), and after ($t=1200$) the jet, viewed from the side (left), above (middle), and along the coronal loop (right). An animation of this figure is available online.}}
\label{fig:bp8LOS}
\end{figure*}

\begin{figure*}[t]
\centering
\includegraphics[width=0.99\textwidth]{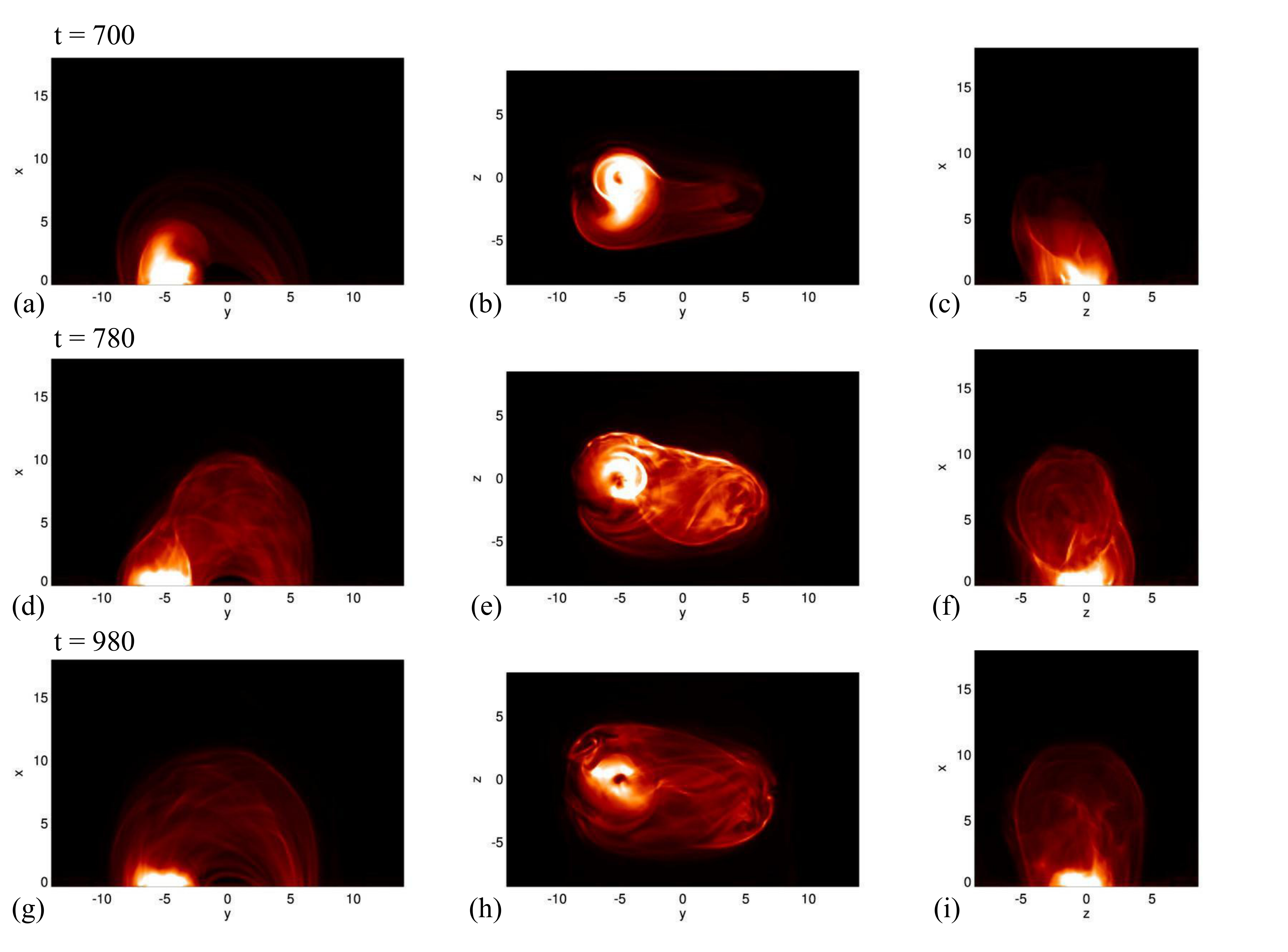}
\caption{{$L/N=1.46$. Line-of-sight integrated current density before ($t = 700$), during ($t = 780$), and after ($t=980$) the jet, viewed from the side (left), above (middle), and along the coronal loop (right). An animation of this figure is available online.}}
\label{fig:bp5LOS}
\end{figure*}

In addition, we expect that there should be heating associated with the filamentary current structures in the loop. The resulting actual EUV or soft X-ray emission is a strongly nonlinear function of the temperature, which in turn is determined by a balance among the local heating and radiative cooling rates plus thermal conduction, none of which is treated by our simulation. For simplicity, we elected to visualize the line-of-sight integral of the current density magnitude  ($\vert {\bf J} \vert$). Similar approaches have been used previously to illustrate the appearance of sigmoids \citep{Archontis2009,Aulanier2005b,Aulanier2010}  and sheared arcades  \citep{Schrijver2008} in various coronal models. Our purpose is to present a qualitative idea of the sort of structures that might be observed generically, not to emulate any particular instrument or to replicate any particular observation.

Figures \ref{fig:bp8LOS} and \ref{fig:bp5LOS} show synthetic images created using this proxy for our jets when viewed from the side, above, and along the connecting coronal loop. For the $L/N=2.40$ jet, filamentary structure is clearly evident in the coronal loop once the jet is underway (Fig.\ \ref{fig:bp8LOS}(d-f)). The sheared-field region beneath the separatrix dome still carries the greatest current and appears as the brightest feature. In the side view, the wandering of the jet spire across the region is seen clearly in the online animation, as the untwisting flux is processed sequentially by the rotating reconnection site. In the post-jet state, the filamentary current layers within the loop appear as criss-crossing streaks that generally follow the twist imparted by the jet  (Figs.\ \ref{fig:bp8LOS}(g-i)). We point out that this proxy for emission captures both the inter-thread current sheets as well as the current-carrying threads themselves. As noted above, {at least some} of the threads {may be} brighter due to their enhanced density. The footpoints of some of the largest flux ropes (corresponding to the largest swirls in the squashing factor evaluated on the photosphere, Fig.\ \ref{fig:qbp8}(d)) also appear as filamentary structures around the base of the separatrix dome (Fig.\ \ref{fig:bp8LOS}(h)). This is even more evident in the relaxing state of the $L/N=1.46$ jet, in the form of a large swirl next to the separatrix when viewed from above (Fig.\ \ref{fig:bp5LOS}(h)). This feature is co-located with a large swirl in the squashing factor (Fig.\ \ref{fig:qbp5}(d)). It {winds up and then unwinds} as it forms and relaxes (see online animation).

The above analysis suggests that in addition to intermittent outflows and blob-like expulsions of plasma, bright filamentary structures in the jet outflows and swirling structures around the base of the jet also are signatures of fragmented reconnection occurring in these events.

\section{Discussion}
\label{sec:discussion}
In this paper, we have analyzed the evolution of tearing-mediated reconnection in two high-resolution numerical simulations of coronal jets. Our jets were at the extremes of the parameter range explored at lower resolution in WD16 and exhibited similar macroscopic behaviors {to those earlier calculations}. By tracking the null points in the volume and analyzing the magnetic connectivity, we were able to pinpoint when tearing began in the simulations and to follow the evolution of the fragmented reconnection region in the jets. 

In agreement with the {idealized 3D null-point reconnection} studies of \citet{Wyper2014a,Wyper2014b}, we find that tearing in the jet current layer leads to the formation of multiple null points and of interacting flux-rope structures. The onset of tearing occurred before the onset of the jet, which resulted from the triggering of a kink-like instability in the twisted field beneath the separatrix dome. The kinking of the twisted field generated favorable conditions for fast reconnection and the rapid release of stored magnetic energy. These dynamics did not occur due to the fragmentation of the current layer. Consequently, the macroscopic behavior of our high-resolution jets is fully consistent with the less well-resolved jets studied by WD16.

The tearing in our jets appears to occur once the current layer becomes sufficiently long and thin, i.e.\ after it reaches a high aspect ratio. We are relying upon numerical resistivity to facilitate reconnection, so the current-layer thickness is set by the grid spacing. At still higher resolution, therefore, tearing might occur even earlier, since any critical aspect ratio can be reached sooner for a fixed sheet length. However, we believe it unlikely that this would substantially alter either the fast reconnection initiated by the kink instability or the subsequent jet generation. To test this requires even more extensive simulations at still higher 3D resolution, which is beyond the scope of this work.

In our jets, the current layer was most fragmented midway through the jet, when the layer reached its most elongated state and the reconnection rate peaked. This tearing generated flux ropes both in the weak-field region at the center of the current layer and along the flanks of the separatrix surface. Using a rough proxy for soft X-ray/EUV emission, we showed that the largest of such flux ropes may be visible as swirls of emission near the base of the jet. We estimate the expected size of such swirls by using their relative size compared to the dome. The largest one that we identified is shown in Figure \ref{fig:bp5LOS}(h) and corresponds to the large swirl in Q in Figure \ref{fig:qbp5}(d). It has a width of around $1$, or about $1/6$ of the width of the dome initially. Assuming a typical dome width of $6~{\rm Mm}$, this swirl has a width of $1 {\rm ~Mm}$, or $1.4$ arc seconds. Such swirls are larger than the limit of resolution with IRIS ($\approx 0.33$ arc seconds) for typical jets, so they may be identifiable in the largest jet events if they appear in cooler chromospheric lines.

Once the flux ropes formed, their inherent twist spread along the field lines as they were ejected from the current layer. They then became torsional wave packets within the main jet outflow. The flux ropes in the weak-field region also had associated density enhancements, forming plasma ``blobs'' that were localized to these structures. Once the flux ropes were ejected, new ones formed in their place, and the process was repeated as the jet proceeded. This repeated formation and ejection of plasma blobs provides a natural explanation for the intermittent outflows, bright blobs of emission, and quasi-periodic intensity fluctuations observed in some jets \citep[e.g.][]{Singh2011,Singh2012,Morton2012,Zhang2014,Filippov2015}. The thread-like nature of the tearing-mediated outflows may also explain the filamentary structure often observed in jets \citep[e.g.][]{Singh2011,Cheung2015}. However, such filamentary structure may also be the result of thermal effects including condensation and evaporation, which are not treated in our simulations. 

In our scenario, the jets were confined along a coronal loop where they transferred twist from beneath the separatrix dome to the larger-scale magnetic field surrounding it. We introduced a simple method for estimating the flux affected by the jet, based on fitting a circle that contains the same amount of flux as the parasitic polarity around the base of the separatrix just prior to the jet. We noted that this method may also be useful for estimating the heliospheric flux affected by coronal-hole jets, along which fast outflows and high-energy particles are expected. The method worked well for us because the outer flux is quite evenly distributed in our simulations. In more realistic fields, where the photospheric flux is distributed in patches, a more complex method involving some form of weighting may be required. 

We also showed that, after a period of relaxation, the final states in each simulation were not simple, uniformly twisted loops as might be expected based on \citet{Taylor1986} relaxation theory. Rather, the loops contained many twisted threads that were remnants of the flux ropes formed by tearing during the jet (Fig.\ \ref{fig:ropeend}). Between the threads were multiple extended current layers that stretched along the loop. This is the 3D version of the reconnection-driven current filamentation described by \citet{Karpen1996}. It is a reminder that coronal reconnection can produce multiple current layers that may heat the coronal loop plasma as they dissipate. Using our rough proxy for emission, we showed that these cooling threads may be observable as criss-crossing thread-like features in the connecting coronal loop. 

{\citet{Sun2013} described a large coronal jet similar in morphology to our $L/N = 2.40$ case. They observed criss-crossing threads within the loop and a two-phase emission suggestive of shuffling reconnection and possible heating within the cooling loop, supporting the picture we have deduced from our simulations. To test this fully requires a more comprehensive treatment of the plasma energetics than the simple model adopted here. Other possible avenues for future research include a realistic stratification of the background atmosphere; an exploration of how particles are accelerated during the jet; where and when jet flows, heat fluxes, and/or particles precipitate at the photosphere to generate remote brightenings; and the resultant occurrence of chromospheric evaporation flows into the corona.}
 
Finally, we note that essentially all jet models invoke reconnection between regions of locally closed and locally open field. Such models implicitly assume that reconnection is occurring at one or more 3D magnetic nulls, so we conclude that this repeated tearing process is likely to occur in all coronal jets. Our schematic Figure \ref{fig:cartoon} of this repeated process in a generic jet scenario could arise due to flux emergence \citep[e.g.][]{Moreno-Insertis2013}, eruption of a mini-filament \citep[e.g.][]{Filippov2015,Sterling2015}, or instability following photospheric twisting \citep[][]{Pariat2009}. As the spatial resolution and temporal cadence of observing instruments increase, it seems inevitable that such structures will be detected increasingly frequently.

\acknowledgments
{This work was supported by P.F.W.'s appointment to the NASA Postdoctoral Program, administered by Oak Ridge Associated Universities and the Universities Space Research Association, and by C.R.D.'s and J.T.K.'s participation with a NASA Living With a Star Focused Science Team on solar jets. The computer resources used to perform the numerical simulations were provided to C.R.D.\ by NASA's High-End Computing program at the NASA Center for Climate Simulation. {We thank the anonymous referee for helpful comments that improved our manuscript.} We are also grateful to David Pontin, Spiro Antiochos, Etienne Pariat, and Kevin Dalmasse for numerous helpful discussions on the topic of solar jets and of 3D magnetic topology. We also thank John Dorelli, Alex Glocer, and Clare Parnell for useful discussions of numerically identifying 3D null points. Several of the figures were created using the Vapor visualization package (www.vapor.ucar.edu).}

\appendix
\section{3D Nulls}
\label{ap:null}
The 3D null points in our simulations were identified using the trilinear method introduced by \citet{Haynes2007}. This method assumes that the magnetic field between the grid points of the numerical domain may be approximated via linear interpolation. After averaging ARMS's face-centered magnetic field values to the cell vertices, we first identify those grid blocks and, subsequently, cells where at least one field component changes sign, a necessary but not sufficient condition for the existence of a 3D null  \citep{Greene1992,Haynes2007}. On each face, a bilinear form for two of the three field components ($B_{1}$ and $B_{2}$) is fitted, and solutions to the quadratic equation for the position of roots $B_{1}=B_{2}=0$ are found \citep{Haynes2007}. For each face with such roots the value of the third component ($B_{3}$) is evaluated at this position. A null exists if for two faces with roots $B_{1}=B_{2}=0$, $B_{3}$ changes sign. In principle, $B_{1}$, $B_{2}$ and $B_{3}$ can be any combination of $B_{x}$, $B_{y}$ and $B_{z}$. We try each combination in turn and designate a cell as containing a null if all three permutations agree. The null position is then found to sub-grid resolution by using a Newton-Raphson routine. The starting position is seeded randomly within the cell and the derivatives in the Jacobian are obtained from the trilinear expansion of each field component. The iteration converges when the following two conditions are satisfied: 
\begin{align}
\Delta x_{nr} \le {1\times 10^{-6}} , \nonumber\\
|\mathbf{B}_{nr}| \le {1\times 10^{-2}}B_{max},
\end{align}
where $\Delta x_{nr}$ is the spatial distance between successive iterations, and $B_{max}$ the maximum absolute value of the magnetic field on the vertices of the cell. If the above are not met within a set number of iterations, or the scheme leaves the cell, the procedure is repeated for another random starting position. If this procedure fails after $10^{7}$ attempts the possible null point is discarded. To measure the isolation of each null the above procedure (excluding the sub-grid resolution step) was repeated for $3\times 3\times 3$ and $5\times 5 \times 5$ boxes around the original cell. We assign orders $1, 2$ or $3$ to nulls identified within just the cell, the $3\times 3\times 3$ box, and the $5\times 5\times 5$ box, respectively. Those nulls with orders of $1$ or $2$ typically have nulls of opposite type in the next cell or two cells over respectively, that act to cancel out the identification of a null in the volume, consistent with the index theorem for 3D nulls \citep{Greene1992}. Thus, nulls of order $1$ are bordered in adjacent cells by a null of opposite type, whereas nulls of order $3$ are isolated by at least two cells in all directions. The type of each null (A or B) is determined by evaluating the eigenvalues of the Jacobian at the null position \citep{Parnell1996}. The derivatives of the Jacobian are obtained from a fourth-order central-difference method applied at the cell vertices and interpolated to the null position. At type A and B nulls, respectively, the magnetic field approaches or recedes from the null along the spine \citep{Lau1990}. We limit the analysis of the nulls to within sub-volumes surrounding the evolving separatrix surface in each simulation. Sub-volumes of $[x,y,z] \in [0.0..12.0,-15.0..-1.0,-7.0..7.0]$ and $[0.0..6.0,-15.0..1.5,-7.0..7.0]$  were used for $L/N=2.40$ and $1.46$, respectively.

\section{Separatrix Surface}
\label{ap:separatrix}
Each new null point formed in our simulation volume must have an associated spine and fan separatrix surface. However, we were interested primarily in the position of the global separatrix surface partitioning the flux connected to the parasitic polarity from that of the coronal loop. When the null region is fragmented, this surface can be formed by the fan planes of some of the multiple null points \citep{Wyper2014b}. We identified the global separatrix by assessing the connectivity of points in a 3D regular grid containing all of the locally closed flux. The grid spacing used was roughly twice the numerical grid spacing at the highest refinement level. At each grid point, field lines were traced forward and backward along the field to find the two footpoints on the photosphere. {The grid point was in the locally open region if one footpoint had $y \ge 0$, i.e. the field line closed down to the photosphere within the far footpoint of the coronal loop.} Closed-field grid points were given a numerical value of $1$ and open-field grid points a value of $0$. The global separatrix surface was then visualized by showing an isosurface of $0.5$. {This method is a simple way to visualize global topological boundaries, assuming the identities of different magnetic domains can be designated by querying the connectivity of the field lines.}

\section{Squashing Factor}
\label{ap:q}
To calculate the squashing factor $Q$ \citep{Titov2002,Titov2007} in our simulations, $1200^2$ field lines were traced from two regular grids centered on the footpoints of the two spine lines. For each grid $Q$ was calculated using
\begin{eqnarray}
Y=Y(y,z) \quad \& \quad Z=Z(y,z), \\
Q = \frac{B_{x}^{*}}{B_{x}} \sqrt{\left(\frac{\partial Y}{\partial y}\right)^{2}+\left(\frac{\partial Y}{\partial z}\right)^{2}+\left(\frac{\partial Z}{\partial y}\right)^{2}+\left(\frac{\partial Z}{\partial z}\right)^{2}}.
\end{eqnarray}
Here $Y$, $Z$ and $B_{x}^{*}$ are the coordinates and component of the field normal to the photosphere at the far ends of the field lines, and $y$, $z$ and $B_{x}$ are the values at the starting positions. The gradients were approximated using a fourth-order central-difference method.

\section{Magnetic Helicity}
\label{ap:helicity}
The relative magnetic helicity H is calculated in the gauge-independent form of Finn \& Antonsen (1985),
\begin{equation}
H = \int_{V}{(\mathbf{A}+\mathbf{A}_{p})\cdot(\mathbf{B}-\mathbf{B}_{p})\,dV},
\end{equation}
where $\mathbf{B}$ is the instantaneous magnetic field with associated vector potential $\mathbf{A}$, $\mathbf{B}_{p}$ is the corresponding current-free (potential) magnetic field having the same normal component as $\mathbf{B}$ at the lower (ÒphotosphericÓ) boundary, $\mathbf{A}_{p}$ is its associated current-free vector potential, and $V$ is the domain volume. Our Cartesian coordinates have $x$ in the vertical direction while $y$ and $z$ span the horizontal plane. Therefore, it is convenient to specify at the lower boundary ($x = x_{0}$) $A_{y}$ and $A_{z}$, which together define $B_{x}$ there, and to select a gauge in which $A_{x} = 0$. In that case,
\begin{equation}
\mathbf{A}(x,y,z,t) = \mathbf{A}_{p}(x=x_{0},y,z,t)+\hat{\mathbf{y}}\int_{x_{0}}^{x}{dx'\,B_{z}(x',y,z,t)}-\hat{\mathbf{z}}\int_{x_{0}}^{x}{dx'\, B_{y}(x',y,z,t)}
\end{equation}
defines $\mathbf{A}$ everywhere in the domain, given the values of $\mathbf{A}_{p}$ at the lower boundary. Our jet simulations are designed to hold $\mathbf{B}_{x}$ fixed at the lower boundary as the system evolves; therefore $\mathbf{B}_{p}$ and $\mathbf{A}_{p}$ are time-independent.

The magnetic field is initialized as the superposition of two sub-surface dipoles: a large-scale horizontal dipole ($\mathbf{A}_{H},\mathbf{B}_{H}$) that provides the background field, and a small-scale vertical dipole ($\mathbf{A}_{V},\mathbf{B}_{V}$) that embeds a local minority-polarity region within the background field and introduces a magnetic null point into the corona above. The potentials are given by
\begin{align}
\mathbf{A}_{V} &= \frac{1}{2}B_{V}d_{V}^{3}\frac{1}{\left[(x-x_{V})^{2}+(y-y_{V})^{2}+(z-z_{V})^{2}\right]^{3/2}}\left[-(z-z_{V})\hat{\mathbf{y}}+(y-y_{V})\hat{\mathbf{z}}\right], \nonumber \\
\mathbf{A}_{H} &= \frac{1}{2}B_{H}d_{H}^{3}\frac{1}{\left[(x-x_{H})^{2}+(y-y_{H})^{2}+(z-z_{H})^{2}\right]^{3/2}} \left[+(z-z_{H})\hat{\mathbf{x}}-(x-x_{H})\hat{\mathbf{z}}\right].
\end{align}
In our simulations we take $(B_{H}, d_{H}, x_{H}, y_{H}, z_{H}) = (8.0,10.0,-10.0,0.0,0.0)$, and $(B_{V}, d_{V}, x_{V}, y_{V}, z_{V}) = (25.0,1.7,-1.7,y_{0},0.0)$. To create field configurations with $L/N=2.40$ and $L/N=1.46$ we set $y_{0}=-8.0$ and $-5$, respectively. $A_{Vx}=0$ fulfills our choice of gauge automatically. To eliminate the $A_{Hx}$ component of the potential, we make the gauge transformation
\begin{equation}
\mathbf{A}'_{H} = \mathbf{A}_{H} - \boldsymbol{\nabla} \psi_{H}.
\end{equation}
The scalar function that cancels $A_{Hx}$ and is well-behaved at $y=y_{H}, z=z_{H}$ is
\begin{equation}
\psi_{H} = \frac{1}{2}B_{H}d_{H}^{3}\frac{z-z_{H}}{(y-y_{H})^{2}+(z-z_{H})^{2}} \left \{ \frac{x-x_{H}}{\left[ (x-x_{H})^{2}+(y-y_{H})^2+(z-z_{H})^{2}\right]^{1/2}} -1\right\}.
\end{equation}
As a result of this transformation, the vector potential
\begin{equation}
\mathbf{A}'_{H} = A'_{Hy} \hat{\mathbf{y}} + A'_{Hz}\hat{\mathbf{z}}
\end{equation}
has the more complicated components
\begin{equation}
A'_{Hy} = \frac{1}{2}B_{H}d_{H}^{3} \frac{(y-y_{H})(z-z_{H})}{(x-x_{H})\left[(x-x_{H})^{2}+(y-y_{H})^{2}+(z-z_{H})^{2}\right]^{3/2}} F(\xi)
\end{equation}
and
\begin{align}
A'_{Hy} =& \frac{1}{2}B_{H}d_{H}^{3} \frac{(z-z_{H})^{2}}{(x-x_{H})\left[(x-x_{H})^{2}+(y-y_{H})^{2}+(z-z_{H})^{2}\right]^{3/2}} F(\xi) \nonumber \\
+& \frac{1}{2}B_{H}d_{H}^{3} \frac{1}{(x-x_{H})\left[(x-x_{H})^{2}+(y-y_{H})^{2}+(z-z_{H})^{2}\right]^{1/2}} G(\xi) \nonumber \\
-& \frac{1}{2}B_{H}d_{H}^{3} \frac{(x-x_{H})}{\left[(x-x_{H})^{2}+(y-y_{H})^{2}+(z-z_{H})^{2}\right]^{3/2}}
\end{align}
where 
\begin{align}
\xi =& \, \frac{(y-y_{H})^{2}+(z-z_{H})^{2}}{(x-x_{H})^{2}}, \nonumber\\
F(\xi) =& \, \frac{1}{\xi^{2}}[2+3\xi-2(1+\xi)^{3/2}], \nonumber \\
G(\xi)=& \, \frac{1}{\xi}[(1+\xi)^{1/2}-1].
\end{align}
Both $F$ and $G$ are finite as their argument $\xi \to 0$.


\end{document}